\documentclass[preprint,secnumarabic,amssymb, nobibnotes, aps, superscriptaddress, prb]{revtex4-1}
\usepackage{graphicx}

\begin{document}

\title{Accelerating the laser-induced demagnetization of a ferromagnetic film by antiferromagnetic order in an adjacent
layer}

\author{I. Kumberg}%
\affiliation{Institut f\"ur Experimentalphysik, Freie Universit\"at Berlin, Arnimallee 14, 14195 Berlin, Germany}
\author{E. Golias}%
\affiliation{Institut f\"ur Experimentalphysik, Freie Universit\"at Berlin, Arnimallee 14, 14195 Berlin, Germany}
\author{N. Pontius}
\affiliation{Helmholtz-Zentrum Berlin f\"ur Materialien und Energie, Albert-Einstein-Stra{\ss}e 15, 12489 Berlin, Germany}
\author{R. Hosseinifar}
\affiliation{Institut f\"ur Experimentalphysik, Freie Universit\"at Berlin, Arnimallee 14, 14195 Berlin, Germany}
\author{K. Frischmuth}
\affiliation{Institut f\"ur Experimentalphysik, Freie Universit\"at Berlin, Arnimallee 14, 14195 Berlin, Germany}
\author{I. Gelen}
\affiliation{Institut f\"ur Experimentalphysik, Freie Universit\"at Berlin, Arnimallee 14, 14195 Berlin, Germany}
\author{T. Shinwari}
\affiliation{Institut f\"ur Experimentalphysik, Freie Universit\"at Berlin, Arnimallee 14, 14195 Berlin, Germany}
\author{S. Thakur}
\affiliation{Institut f\"ur Experimentalphysik, Freie Universit\"at Berlin, Arnimallee 14, 14195 Berlin, Germany}
\author{C. Sch\"u{\ss}ler-Langeheine}
\affiliation{Helmholtz-Zentrum Berlin f\"ur Materialien und Energie, Albert-Einstein-Stra{\ss}e 15, 12489 Berlin, Germany}
\author{P.~M. Oppeneer}
\affiliation{Department of Physics and Astronomy, Uppsala University, Box 516, SE-75120, Uppsala, Sweden}
\author{W. Kuch}
\affiliation{Institut f\"ur Experimentalphysik, Freie Universit\"at Berlin, Arnimallee 14, 14195 Berlin, Germany}

\date{\today}%

\begin{abstract}
	We study the ultrafast demagnetization of Ni/NiMn and Co/NiMn ferromagnetic/antiferromagnetic bilayer systems after excitation by a laser pulse. We probe the ferromagnetic order of Ni and Co using magnetic circular dichroism in time-resolved pump--probe resonant X-ray reflectivity.  Tuning the sample temperature across the antiferromagnetic ordering temperature of the NiMn layer allows to investigate effects induced by the magnetic order of the latter.  The presence of antiferromagnetic order in NiMn speeds up the demagnetization of the ferromagnetic layer, which is attributed to bidirectional laser-induced superdiffusive spin currents between the ferromagnetic and the antiferromagnetic layer.  

\end{abstract}
\maketitle

\section{Introduction}
Magnetic recording and storage media will face major challenges in the near future regarding energy efficiency and data transfer speed due to increased demand and continuing growth of data storage volume. Much could be gained by switching from field-induced manipulation of magnetic order, where speed is limited by spin precession dynamics \cite{Back.1998, Schumacher.2003, Tudosa.2004}, to manipulation by ultra short light pulses
\cite{Beaurepaire.1996, Koopmans.2005, Cinchetti.2006, Bigot.2009, Koopmans.2010, Kirilyuk.2010, Lambert.2014, Walowski.2016}. 
To explain ultrafast magnetization dynamics following the excitation by an ultrashort laser pulse, a number of different approaches are being discussed.  One can distinguish between local mechanisms such as scattering with (quasi-)particles \cite{Koopmans.2005,Cinchetti.2006,Stamm.2007,Krau.2009,Turgut.2016} and non-local mechanisms such as transport by superdiffusive spin currents \cite{Battiato.2010,Eschenlohr.2013}. In heterostructures, the latter may dominate the observed dynamic behavior since they allow transfer of angular momentum between different ferromagnetic layers, which may even lead to an entirely different dynamic behavior as compared to isolated layers \cite{Rudolf.2012,Turgut.2013}.  Transient superdiffusive spin currents are also expected to occur at the interface between an antiferromagnet (AFM) and a ferromagnet (FM).  Although unpolarized in the interior of an AFM layer, superdiffusive currents excited by a laser pulse become spin-polarized in the vicinity of the interface to an FM layer because of spin-polarized interface reflection and transmission.  In this work, we present a study on the ultrafast optical demagnetization of an FM/AFM layered system. We show that the presence of antiferromagnetic order in an adjacent layer accelerates the demagnetization of the FM layer, which we attribute to the exchange of superdiffusive spin currents between the two layers.

FM/AFM layered systems are well known in device design, as the exchange interaction between the two layers leads to an exchange bias (EB) effect, resulting in the magnetic pinning of the FM layer \cite{Meiklejohn.1956, Nogues.1999}. 
NiFe/NiO as an FM/AFM system has been already studied by time-resolved magneto-optical Kerr effect in the pioneering works of Ju \textit{et al}. where an unpinning of the interface spins within times $\leq 1$ ps was reported \cite{Ju.1998,Ju.1999,Ju.2000}., More recently, the torque on the FM spins in an FM/AFM bilayer after laser excitation was investigated \cite{DallaLonga.2010}, suggesting, like the works of Ju \textit{et al}., that the magnetization in such a system is switchable by a laser pulse.
In extension of these previous investigations, which were mainly targeting the evolution of EB after laser excitation and the correlated precessional motion of the FM macrospin, we investigate here the ultrafast processes in an FM/AFM bilayer system after an optical excitation pulse. Using Ni and Co as ferromagnets and NiMn as an antiferromagnet, we have grown epitaxial FM/AFM bilayer samples with easily accessible N{\'e}el temperature $T_N$ of the AFM layer.  We employ resonant X-ray magnetic circular dichroism in reflectivity to probe the magnetization with elemental resolution. Comparing the demagnetization behavior above and below $T_N$ allows us then to investigate the influence of the magnetic order in the NiMn layer on the demagnetization of Ni and Co following the excitation with a femtosecond laser pulse. We observe a significantly faster demagnetization for the same amount of demagnetization of the FM layer if the adjacent NiMn layer is antiferromagnetically ordered. We discuss this in terms of the exchange of angular momentum through the FM/AFM interface by superdiffusive spin currents.

\section{Experiment} 
The samples are grown by molecular beam epitaxy in our home lab at the Freie Universit{\"a}t Berlin. All films are prepared on Cu(001) single crystals, which were cleaned by multiple Ar$^+$ sputtering and annealing cycles. The surface integrity is verified by LEED and the sample's cleanliness by Auger electron spectroscopy. The materials are thermally evaporated by electron bombardment from a rod of 99.998\% purity for Ni and Co, and from Mn flakes with 99.98\% purity in a Ta crucible. During evaporation the sample is kept at room temperature and deposition rates of about 1 monolayer (ML) per minute are used. The pressure is kept below $1\times10^{-9}$ mbar while evaporating, with a base pressure of  $\approx 6 \times 10^{-10}$ mbar. Deposition is monitored by medium-energy electron diffraction to ensure layer-by-layer growth. Composition and thickness are additionally checked and calibrated by the signal intensity in Auger electron spectroscopy. After preparation, the samples are capped by 20 ML Cu with purity of 99.99\%, evaporated from a Ta crucible, to prevent sample oxidation during transport. We prepared a 20 ML Cu /15 ML Co /20 ML Ni$_{31}$Mn$_{69}$/Cu(001) and a 20 ML Cu /12 ML Ni /14 ML Ni$_{38}$Mn$_{62}$/Cu(001) sample. For the AFM layer we chose NiMn, an alloy which orders antiferromagnetically around equiatomic concentrations. Furthermore, EB can occur both in in-plane and out-of-plane geometries \cite{Tieg.2006,Hagelschuer.2016} and a vanishing of direct exchange coupling between out-of-plane-magnetized FM layers across NiMn indicates a spin reorientation, possibly between an in-plane and an out-of-plane AFM spin structure at temperatures about halfway between the N\'{e}el temperature and the onset of EB \cite{Hagelschuer.2016}. For the present investigation we chose a stoichiometry close to Ni$_{40}$Mn$_{60}$, the Ni to Mn ratio for which we have a detailed investigation of the N\'{e}el-Temperature \cite{Hagelschuer.2016}. Both, the thickness and the concentration, were chosen to have T$_N$ around 360 K \cite{Hagelschuer.2016,Tieg.2006,Reinhardt.2010}. For the FM layer we selected Ni with an out-of-plane and Co with an in-plane easy axis of magnetization.
 
The experiments are carried out at the femtoslicing facility at beamline UE56/1 ZPM of the synchrotron radiation source BESSY II in Berlin.  This facility provides ultra-fast soft X-ray pulses for time-resolved experiments \cite{Holldack.2014}. The setup allows to measure the transmitted or reflected intensity of the circularly polarized X-ray probe pulse following the pump laser pulse. The system measures at 6 kHz repetition rate, alternating between X-ray probe pulses with and without a leading laser pump pulse with variable pump--probe delay time. This measurement scheme allows for quasi-simultaneous recording of the pumped and the unpumped signal.  The X-ray and laser spot sizes are 140 $\mu$m $\times$ 40 $\mu$m and 1500 $\mu$m $\times$ 200 $\mu$m, respectively. Both are co-propagating with an angle of $1-2^{\circ}$ between them to separate them again after reflection at the sample. By switching the magnetization of the sample by an external magnetic field, we make use of the X-ray magnetic circular dichroism (XMCD) in reflection \cite{Mertins.2002} as an element-resolved probe of the magnetization. All measurements are performed in saturation conditions under applied magnetic fields of $\pm 300$ mT for Co/NiMn and $\pm 400$ mT for Ni/NiMn. The laser pulse has a temporal full width at half maximum (FWHM) of 60 fs and the X-ray pulse of 100 fs, with a resolving power of E/$\Delta$E = 500.

 \begin{figure}
	\includegraphics[width = 1\linewidth]{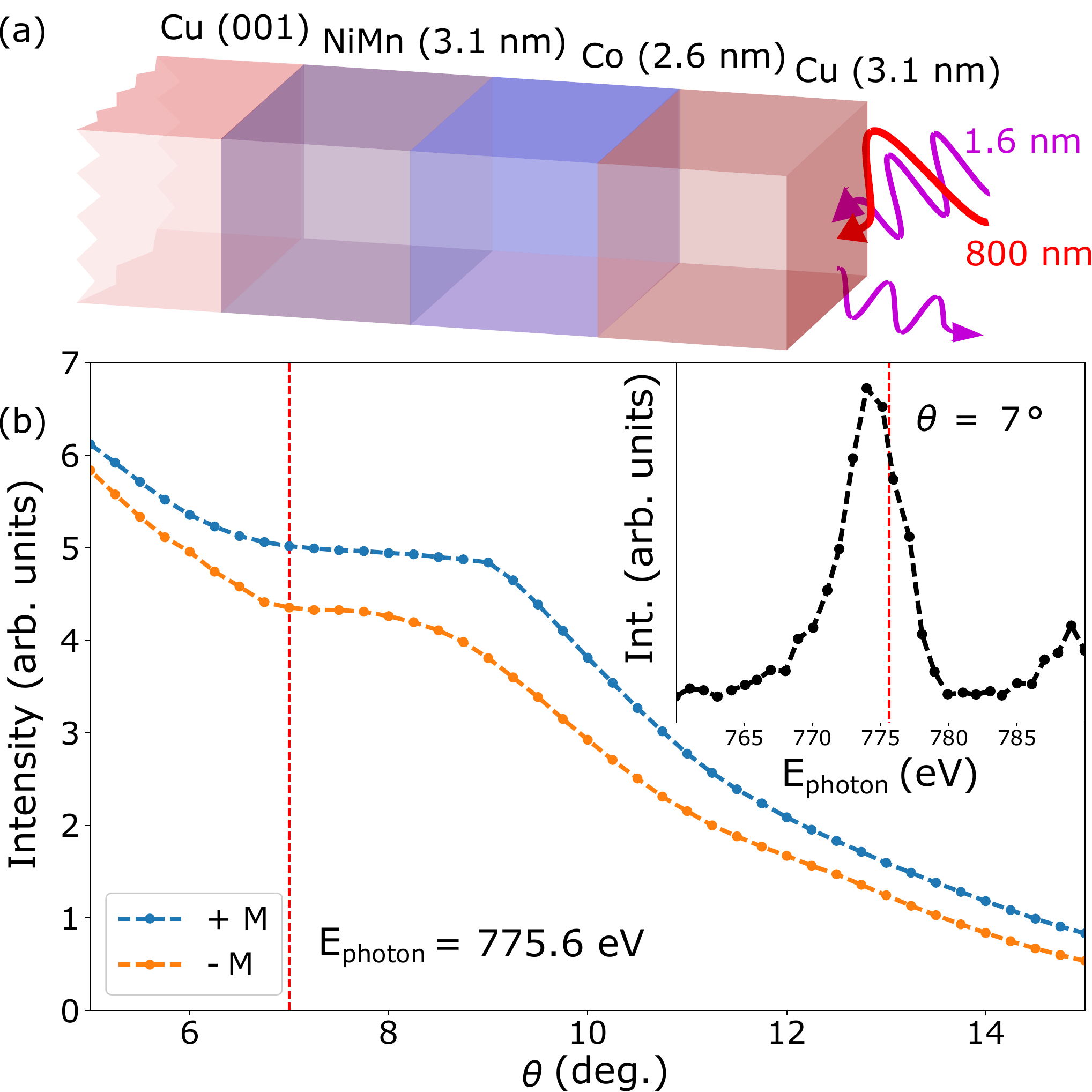}
	\caption{(a) Schematic representation of the Co/NiMn sample. The measurements are performed in reflectivity with an 800 nm pump and 775.6 eV ($\approx 1.6$ nm) probe pulse. 
		(b) Optimum measurement conditions are found by taking spectra at different angles and comparing $\theta-2\theta$ scans at different photon energies. The most efficient conditions with respect to a minimized data acquisition time for Co are found at $\theta = 7^{\circ}$, highlighted by a vertical dashed line in the figure. The optimum photon energy for Co is identified as $E = 775.6$ eV, highlighted by the dashed vertical line shown in the inset of the figure, where a spectrum recorded at $7^{\circ}$ incidence is shown. }
	\label{spectrum and tth}
\end{figure} 

To find the optimum measurement conditions, a range of static reflectivity scans are performed, identifying the angle with the highest magnetic contrast for a given photon energy while still providing enough photon intensity for the slicing measurements. Energies around the $L_3$ edges of Ni and Co were checked and chosen such that the anticipated data acquisition time is minimized. For Co the optimum conditions are found at $7^{\circ}$ angle of incidence with respect to the surface and 775.6 eV photon energy, indicated in the reflectivity and energy scans presented in Fig.\ \ref{spectrum and tth}. In a similar fashion, the optimum conditions for Ni are identified at $6.5^{\circ}$ angle of incidence and a photon energy of 845.0 eV. 

\begin{figure}
	\includegraphics[width = 1\linewidth]{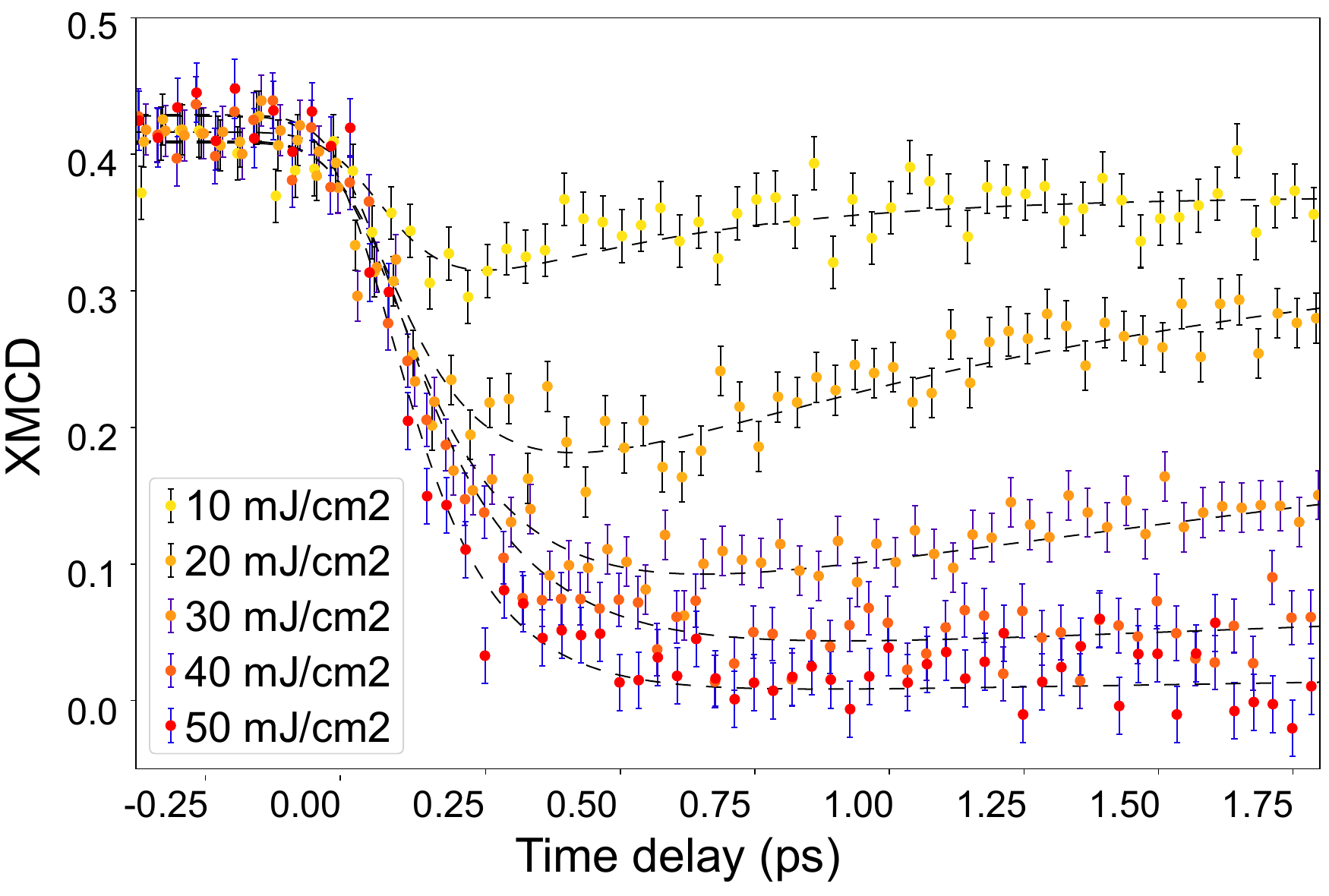}
	\caption{Time delay traces of the Co demagnetization in Co/NiMn. Here the XMCD at the Co $L_3$ edge 
	as a function of the delay at 390 K for different laser fluences is shown. With increased laser fluence, it takes longer to reach the minimum magnetization, and also the remagnetization takes longer. Complete demagnetization is reached with 50 mJ/cm$^2$ incident laser fluence.
	}
	\label{Cohigh T}
\end{figure}

The range of pump fluences investigated is chosen such that the system completely demagnetizes at the highest fluence and shows a considerable demagnetization at the lower end, in order to study the influence of different excitation strengths on the system. The cobalt system is thus investigated with incident fluences between 10 and 50 mJ/cm$^2$, covering a demagnetization range from 20 to 100\%, as shown in Fig.\ \ref{Cohigh T}.
For Ni/NiMn we recorded data for two different incident fluences of 15 and 46 mJ/cm$^2$, resulting in 60 and 90\% relative demagnetization, respectively. From a layerwise absorption calculation the surface-transmitted fluence is estimated to be about 15-20 \% of the incident fluence, see appendix.  
To investigate the effect of the magnetic order of the antiferromagnetic layer on the demagnetization dynamics, two base temperatures of 80 and 390 K for Co/NiMn and 80 and 360 K for Ni/NiMn were chosen, respectively. While the FM layer keeps its magnetization, the NiMn layer undergoes a phase transition from AFM at low temperatures to paramagnetic at the higher base temperature.

\section{Results}

\begin{figure}[ht]
	\includegraphics[width = 1.\linewidth]{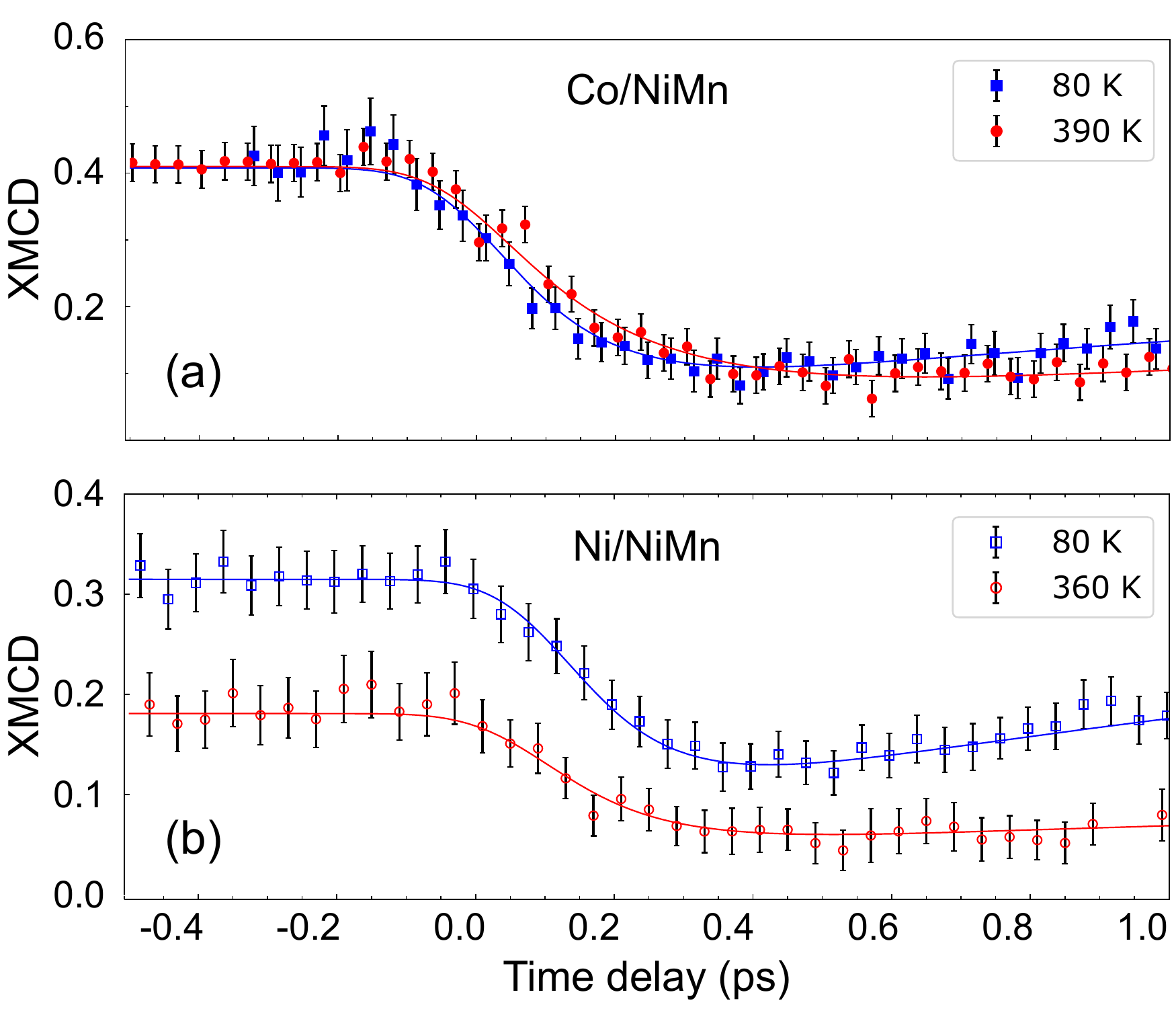}
	\caption{Selected time delay traces for the demagnetization of Co and Ni in Co/NiMn and Ni/NiMn, respectively. Red circles indicate the high and blue squares the low base temperature, as indicated in the legends; the filled symbols
		show data for the Co/NiMn sample and open symbols for the Ni/NiMn sample. Panel (a) shows delay traces of the demagnetization of Co at 30 mJ/cm$^2$ laser fluence at 80 and 390 K, panel (b) the demagnetization of Ni/NiMn after laser excitation with 15 mJ/cm$^2$ laser fluence.}
	\label{CoNi vert}
\end{figure}

Fig.\ \ref{Cohigh T} shows delay traces of the time-dependent XMCD of the Co/NiMn sample, measured at the Co $L_3$ edge, for 390 K sample temperature at different incident pump fluences. The XMCD is defined as the asymmetry of the intensity of the reflected beam with respect to the direction of the applied magnetic field.
Upon excitation by the laser pulse, a decrease of the XMCD signal to a minimum in XMCD before 600 fs is observed in all cases. With increasing fluence the minimum shifts to longer times and the total change in XMCD increases, following the expected behavior for ferromagnetic thin films after a higher heat load \cite{Koopmans.2010}. In order to compare the dynamics in Ni/NiMn and Co/NiMn, two measurements at different fluences that show about the same relative demagnetization are presented in Fig.\ \ref{CoNi vert}.
In the Co/NiMn sample [Fig.\ \ref{CoNi vert} (a)], the XMCD before time zero is virtually unchanged upon increase of the sample temperature due to the relatively high T$_C$ of Co $\sim 1300$ K at 15 ML thickness \cite{Schneider.1990}, decreasing only from 0.42 to 0.41. This is in contrast to the Ni/NiMn sample, where the XMCD decreases from 0.32 at 80 K to 0.18 at 360 K, shown in Fig.\ \ref{CoNi vert} (b) at negative delay times. 

While the static measurement does not reveal a difference between the two temperatures for Co/NiMn, the dynamic XMCD shows a different time evolution.
The decrease of magnetization of the Co layer at the lower base temperature, where the NiMn layer is antiferromagnetically ordered, is distinctly faster compared to the higher base temperature.  To evaluate this quantitatively, all delay traces recorded were fitted using the sum of two exponentials describing a fast demagnetization and a subsequent slower initial remagnetization, respectively, from which the de- and remagnetization time constants are evaluated. The sum of the exponentials is convoluted with a Gaussian function of FWHM = 120 fs to account for the experimental resolution. The fit function corresponds thus to
\begin{eqnarray}
\!\!\!\! \frac{M}{M_0}(t) &=&  g(t) \otimes \Big( C  \nonumber \\
 &+&  \! \left[ a (e^{-\frac{t-t_0}{\tau_{d}}}-1)-b (e^{-\frac{t-t_0}{\tau_{r}}}-1) \right] \! \theta (t-t_0) \! \Big) ,
\end{eqnarray}  
with the initial XMCD $C$, the amplitudes for de- and remagnetization $a$ and $b$, the de- and remagnetization time constants $\tau_{d}$ and $\tau_{r}$, the Gaussian profile $g(t)$ of the instrumental resolution, and the Heaviside step function $\theta$ at $t_0$. The curves are fitted with different amplitudes for de- and remagnetization, since the complete remagnetization dynamics may not be adequately described by a single exponential. However, for the time window 2 ps used for the fitting, the remagnetization, which is anyways not in the focus of the present investigation, can be well reproduced with a single exponential. As the relative timing of pump and probe pulses shows slow drifts over the time period of this study, the curves plotted in this work are shifted by $t_0$ obtained from the fits, in order to have a common zero for all graphs.

\begin{figure}[ht]
	\includegraphics[width = 0.9\linewidth]{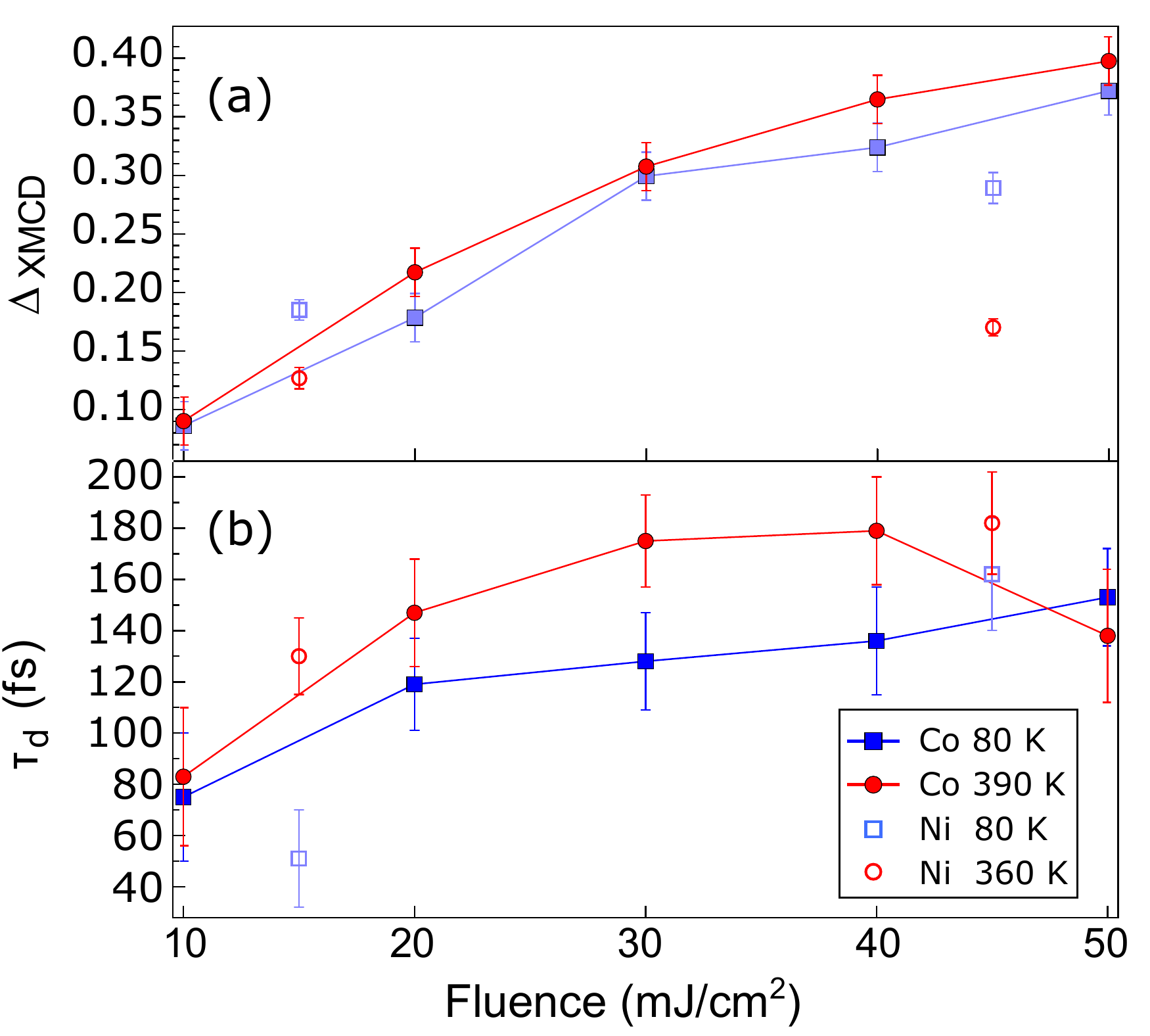}
	\caption{(a) Demagnetization amplitude in XMCD asymmetry after laser excitation for Co (filled symbols) and Ni (open symbols), at 80 K and 360 K for Ni and 390 K for Co, respectively, as denoted in the legend in panel (b). The amplitude of the demagnetization follows a linear trend at lower fluences and is slightly higher for elevated sample temperatures for cobalt, considering the same fluence.
	(b) Magnetization decay time $\tau_d$ versus fluence. Generally, a higher base temperature leads to longer demagnetization times with the largest difference for Co/NiMn around 30 to 40 mJ/cm$^2$ laser fluence.}
	\label{Q and T vs Fluence}
\end{figure}
A fluence-dependent comparison of the demagnetization times and amplitudes for different temperatures is presented in Fig.\  \ref{Q and T vs Fluence}.  For both base temperatures and samples, the quench of the magnetization, depicted in Fig.\ \ref{Q and T vs Fluence} (a) as change of XMCD $\Delta_{XMCD}$ in the same absolute values as on the vertical axes in Figs.\ \ref{Cohigh T} and \ref{CoNi vert}, increases with fluence. Except for the highest fluence, this is true also for the demagnetization time $\tau_{d}$. For the higher sample temperature, demagnetization times are longer and the relative demagnetization stronger, similar to the observations in Ref.\  \cite{Roth.2012}. In absolute values, i.e., in XMCD asymmetry $\Delta_{XMCD}$, the cold Ni/NiMn sample demagnetizes more than the warm one, but when considering the relative amount of magnetization quenched at the lower temperature it is actually less than at the higher temperature. When the Co/NiMn system fully demagnetizes, which is the case for 50 mJ/cm$^2$ at 390 K, the demagnetization is again faster.

\begin{figure}[ht]
	\includegraphics[width = 1.\linewidth]{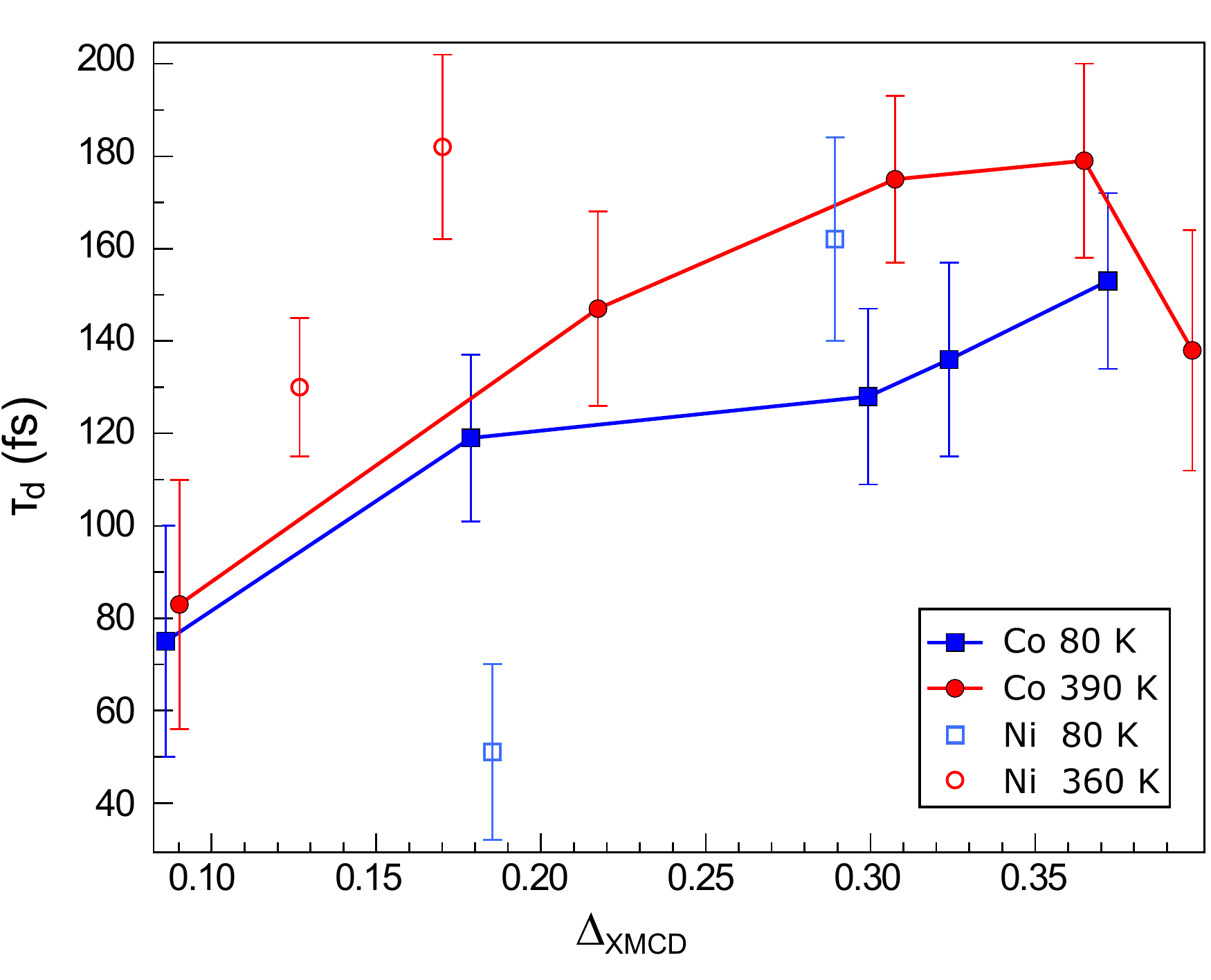}
	\caption{Demagnetization times $\tau_d$ vs.\ demagnetization amplitude for Co/NiMn (filled symbols) and Ni/NiMn (open symbols) at 80 K (blue) and 360 K for Ni and 390 K for Co, respectively (red). The demagnetization time, considering a similar quench of XMCD asymmetry, shows a difference for the two temperatures measured. In Co/NiMn, for both base temperatures an XMCD of 0.42 for 80 K and 0.41 for 390 K was measured before time zero, while for Ni/NiMn an XMCD asymmetry of 0.32 for 80 K and 0.18 for 360 K was observed at negative delay times.
	}
	\label{T vs Q}
\end{figure}

To obtain more information on the role of the antiferromagnetic order in the NiMn layer, we compare the magnetization decay times for Co and Ni as a function of the reduction of the XMCD signal in Fig.\ \ref{T vs Q}. For Co/NiMn, the fluence range covers similar demagnetization at both temperatures, while for Ni/NiMn, where the static difference in magnetization is already quite sizeable, this is not the case.
We observe a difference of $\tau_{d}$ of up to 50 fs for Co/NiMn for the two temperatures around a demagnetization corresponding to a change of the XMCD of 0.3. The difference, albeit small, is still greater than the experimental error and the uncertainty from fitting.  The data for Ni/NiMn suggest that the difference in demagnetization times for demagnetization amplitudes around 0.15--0.20 could be even greater than that.

\section{Discussion}

The comparison of the demagnetization time constants at different temperatures for equal demagnetization amplitudes (Fig.\ \ref{T vs Q}) highlights the role of the AFM order in the NiMn layer for the speed of demagnetization of the FM layer.  
The results show that the demagnetization rate of the FM layer is higher if the AFM layer exhibits magnetic order. It makes the inclusion of new dissipation channels introduced by the adjacent AFM layer necessary for the description of the demagnetization dynamics. 
We suggest in the following that superdiffusive spin currents \cite{Battiato.2010, Rudolf.2012, Eschenlohr.2013, Turgut.2013} are responsible for the faster demagnetization of the FM layer at the lower temperature, i.e., in the presence of antiferromagnetic order in the NiMn layer.  

Transport by superdiffusive currents plays an important role for the ultrafast demagnetization as shown, for example, in Refs.\ \cite{Rudolf.2012, Eschenlohr.2013, Turgut.2013}, where a contribution of transport to an adjacent layer to the ultrafast demagnetization has been reported.  
Before we discuss superdiffusive spin currents in the sample, we take a short look at the simulated differential absorbed laser fluence across the Co/NiMn bilayer, presented in the Appendix, Fig.\ \ref{diffabs}. The energy absorbed in the Co layer is about the same as the one absorbed in the NiMn layer, with a difference of only about 1--2\%.  Hot electrons will thus be excited both in the Co and in the NiMn layer. Excitations in the FM Co layer give rise to a superdiffusive spin current that enters the NiMn layer as well as the cap layer. Similarly, the excitation of the NiMn layer will lead to superdiffusive hot electrons that will be transported through the interface into the FM layer and the Cu substrate. We will discuss in the following how these superdiffusive spin currents contribute to the observed difference in FM-layer demagnetization times below and above the AFM ordering temperature.  

The decisive aspect for the demagnetization of the FM layer is the dissipation of angular momentum by the exchange of spin currents with the NiMn layer. 
Although spin mixing is enhanced after pumping, the superdiffusive spin current carried by hot electrons from the FM layer consists mainly of spin-majority species \cite{Battiato.2010}. Below $T_N$, we assume the magnetization in the FM layer to be collinear with the magnetization axis of the sublattices of the ordered AFM NiMn layer. This configuration results in an injection of the spin-polarized electrons into empty states of the AFM of the same spin character. In the high-temperature paramagnetic state, though, the magnetic moments in the NiMn layer are disordered, resulting in a spin-character mismatch due to the random relative orientation of the spin moments in the NiMn layer with respect to the FM-aligned spins in Co. Under such conditions, the spin penetration depth becomes strongly reduced to about one nanometer \cite{Ghosh.2012}, leading to reflection of carriers at the interface.
Conversely, spin-polarized electron penetration into magnetic systems with collinear spin orientation is much larger 
\cite{Alekhin.2017}. 
In the AFM-ordered state with a collinear spin structure, the superdiffusive majority-type spin current from the FM layer can thus propagate much farther into the NiMn layer, making it a more efficient sink for angular momentum. This leads to shorter demagnetization times for the FM layer adjacent to the ordered AFM layer, as is observed experimentally.

In turn, the superdiffusive current of hot electrons originating in the NiMn layer will be transported through the interface into the FM layer. Likewise, for collinear AFM spin order in the NiMn layer, interface reflection is reduced compared to the case of a magnetically disordered NiMn layer. Despite the absence of a macroscopic magnetization in an AFM, a spin current representing the minority species of the FM layer will be preferentially entering the FM layer, as there are more unoccupied minority-spin states available, thus also accelerating its demagnetization.  

We now argue that the change of temperature alone, without considering the ordering transition in the NiMn layer, cannot explain our experimental observation.  The effect of temperature has not yet been discussed in detail in the model of superdiffusive transport (see \cite{Battiato.2010} and \cite{Battiato.2012}), where the FM layer is considered fully magnetic. In general, spin lifetimes and spin-dependent velocities of the excited electrons change with temperature. This could play a role in the Ni/NiMn sample, where the asymmetry in the spin lifetimes and spin velocities will be reduced at 360 K, in the same way as the static XMCD. For Co, though, a change of this asymmetry is negligible since the change of the experimentally measured XMCD is marginal between the two temperatures.

In addition, the excited electron lifetime decreases at higher temperatures due to enhanced scattering probabilities. Furthermore, slightly different states will be addressed at different temperatures, depending on the slope of the band structure.  This could influence the electron velocity in either direction, but this is also considered a negligible effect here as the change in temperature is only about 300 K. 
Following this line of argumentation, we do not expect a significant change in the demagnetization time for equal demagnetization as a result of the different base temperature in the FM.

	Next we exclude local mechanisms for angular momentum dissipation as a possible explanation. Local demagnetization can be caused by Elliott-Yafet electron--phonon spin-flip scattering \cite{Koopmans.2010} or ultrafast magnon generation \cite{Carpene.2008, Turgut.2016, Eich.2017, Gort.2018}. In the work of Schellekens \textit{et al}.\ \cite{Schellekens.2013}, local demagnetization due to Elliott-Yafet spin-flip scattering has been proposed as the main driving mechanism that could be orders of magnitude stronger compared to spin transport. However, the relative contributions of the microscopic processes strongly depend on the system and properties under investigation \cite{Wieczorek.2015,Turgut.2016}. While electronic-structure-based calculations of the Elliott-Yafet electron--phonon spin-flip scattering predict that this can only give a small contribution to the laser-induced demagnetization \cite{Carva.2011,Essert.2011}, others suggested that a feedback mechanism of the magnetization change and the spin-dependent chemical potential on the band structure would lead to an additional reduction of the exchange splitting \cite{Schellekens.2013b,Mueller.2013}. 
	 However, such a collapse of the exchange splitting has not been found in
	femtosecond spin-polarized photoemission experiments \cite{Eich.2017} and also not in extreme
	ultraviolet transversal magneto-optical Kerr effect (TMOKE) measurements \cite{Turgut.2016} on
	Co.  In any case, using the expressions given in Ref. \cite{Carva.2011} one can calculate that
	an increase in temperature from 80 to 390 K leads to an \textit{increase} in the
	Elliott-Yafet spin-flip scattering by about a factor of 2.5 due to the increased
	phonon population.  This corresponds to a faster electron--phonon demagnetization
	of Co at higher temperatures, which is, however, opposite to the here-measured
	trend, which shows a \textit{slower} demagnetization of Co in Co/NiMn at 390 K.

Magnetization dissipation by transversal spin excitations, i.e., ultrafast magnon generation \cite{Turgut.2016,Eich.2017,Gort.2018}, is another mechanism we exclude, based on the timescale of our observations. Analogous to the case of ultrafast magnon generation followed by spin pumping into an adjacent layer \cite{Choi.2014,Choi.2014b}, 
magnons excited in the FM layer by ultrafast carriers could couple to magnon modes in the ordered AFM layer, a coupling that would be absent in the disordered case. This could also lead to a faster demagnetization of the FM layer when in contact with the ordered AFM layer. However, the timescale of the initial energy transfer process from hot electrons to hot magnons in an FM has not been definitely established. Effects of hot magnon excitation have been detected in laser-excited Co at already 100 fs \cite{Eich.2017}.  The timescale to couple the hot FM magnons to magnon modes in the AFM is neither precisely known. Magnon transport, assuming even a quite high magnon group velocity of $\sim$10$^4$ m/s \cite{Wu.2018}, would only result in an angular momentum transfer over 1 nm in 100 fs, which is not enough to explain the faster demagnetization timescale seen in Fig.\  \ref{T vs Q}.  Thus, although the angular momentum transfer process, starting with hot magnons in the FM layer and their subsequent diffusion into the adjacent layer, is possible, it is expected to occur on a longer timescale than transfer by superdiffusive spin currents.

We finally note that, although our data from the Ni/NiMn sample do not allow the comparison of demagnetization times for the two temperatures at equal demagnetization amplitudes, we expect the same effect as in the Co/NiMn sample.  This is motivated by the Ni data presented in Fig.\ \ref{T vs Q}, where one pair of data points of almost equal magnetization quench shows a much faster rate at the lower temperature.

\section{Conclusions and Outlook}
In conclusion, we infer from our experimental findings superdiffusive spin transport to play a significant role for the demagnetization of an FM/AFM system. A general temperature dependence of demagnetization as well as the contribution of Elliott-Yafet processes cannot account for the observed temperature dependence of the demagnetization. Moreover, a magnon-related contribution to the demagnetization is expected to occur on a longer timescale. Therefore, we attribute the accelerated demagnetization at low temperatures in Co/NiMn and Ni/NiMn to the presence of AFM order in the NiMn layer.

Superdiffusive spin currents can be exchanged more easily between the FM layer and the adjacent AFM-ordered NiMn layer, while the disordered NiMn layer acts as a barrier for the penetration of superdiffusive currents. The AFM-ordered NiMn layer thus represents on the one hand a  sink for majority spin current from the FM layer, while on the other hand it injects minority spin current into the FM layer. Both accelerates the FM-layer demagnetization and explains the experimental observations.  

To accelerate the ultrafast demagnetization by AFM spin order in an adjacent layer may become important for the ultrafast optical manipulation of magnetic order in magnetic recording or spintronic devices. Rotating the spin axis of the AFM layer between perpendicular and parallel to the FM-layer magnetization, for example by staggered spin-orbit torques \cite{Wadley.2016}, could be a means of tuning the demagnetization speed of an FM layer.  
It might finally be interesting to see, possibly by time-resolved magnetic linear dichroism, how in a reverse experiment the presence of FM spin order in an adjacent layer influences the optical quench of the magnetic order in an AFM layer.

\begin{acknowledgments}
This work was supported by the Deutsche Forschungsgemeinschaft via the CRC/TRR 227 ``Ultrafast Spin Dynamics", projects A03 and A07, and by the Swedish Research Council (VR).  We further acknowledge support from the K.\ and A.\ Wallenberg Foundation (grant No.\ 2015.0060) and the Swedish National Infrastructure for Computing (SNIC).
We thank the Helmholtz-Zentrum Berlin for the allocation of synchrotron radiation beamtime.
\end{acknowledgments}

\appendix
\section{Differential Absorption}
 \begin{figure}[ht]
	\includegraphics[width = 0.9\linewidth]{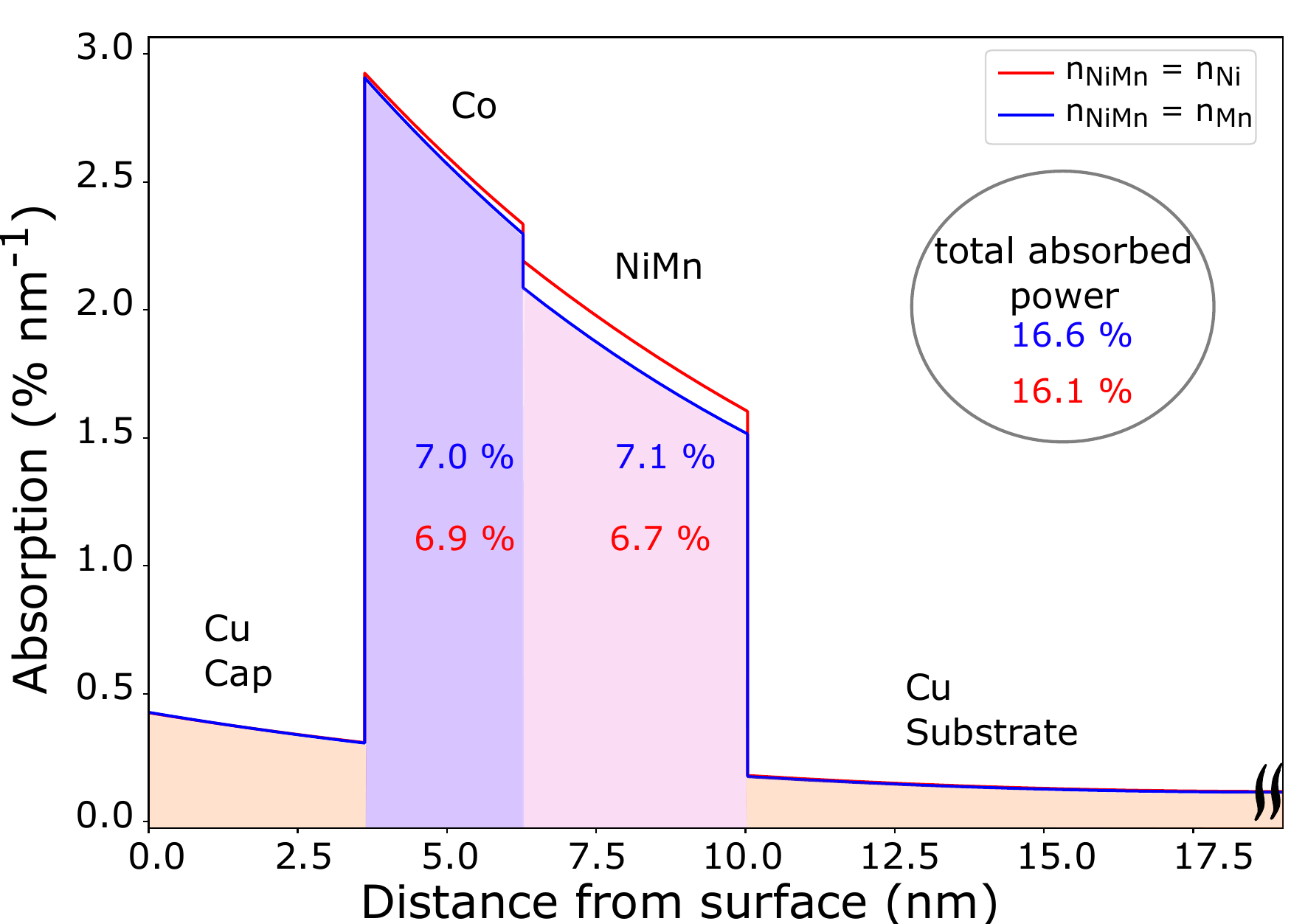}
	\caption{Differential absorption over sample depth for the Co/NiMn sample. The optical constants for NiMn are estimated to lie between the ones for bulk Ni and for bulk Mn. To estimate the absorbed energy in each layer, two absorption profiles are shown, one with the optical constants of Ni for the NiMn alloy (red) and one with the optical constants of Mn (blue). 
		The numbers indicate the fraction of absorbed pump-pulse fluence in each layer.}
	\label{diffabs}
\end{figure}
	
We simulate the layerwise absorption of the pump pulse in the Co/NiMn sample using the matrix formalism described in Ref.\ \cite{Ohta.90}.  The optical constants of NiMn alloy are approximated by calculating the boundary scenarios of a pure Ni or Mn film. To estimate the absorption, both cases are presented in Fig. \ref{diffabs}, where the NiMn alloy absorption is calculated once with the constants of Ni as $n_{Ni} = 2.2180 + i\, 4.8925$ and once with the ones of Mn, $n_{Mn} = 2.7880 + i\, 3.9982$, both taken from Ref.\ \cite{Johnson.1974}. The vertical layer distance for NiMn has been measured at the present concentration as $1.88$ \AA/ML \cite{Tieg.2006}, for Co we use 1.74 \AA/ML \cite{Cerda.1993} and for Cu $1.81$ \AA/ML. In both cases the absorption in the Co layer is slightly higher, but each layer absorbs roughly the same magnitude, $\approx 7$\%, of the incident fluence.


\begin{thebibliography}{50}%
	\makeatletter
	\providecommand \@ifxundefined [1]{%
		\@ifx{#1\undefined}
	}%
	\providecommand \@ifnum [1]{%
		\ifnum #1\expandafter \@firstoftwo
		\else \expandafter \@secondoftwo
		\fi
	}%
	\providecommand \@ifx [1]{%
		\ifx #1\expandafter \@firstoftwo
		\else \expandafter \@secondoftwo
		\fi
	}%
	\providecommand \natexlab [1]{#1}%
	\providecommand \enquote  [1]{``#1''}%
	\providecommand \bibnamefont  [1]{#1}%
	\providecommand \bibfnamefont [1]{#1}%
	\providecommand \citenamefont [1]{#1}%
	\providecommand \href@noop [0]{\@secondoftwo}%
	\providecommand \href [0]{\begingroup \@sanitize@url \@href}%
	\providecommand \@href[1]{\@@startlink{#1}\@@href}%
	\providecommand \@@href[1]{\endgroup#1\@@endlink}%
	\providecommand \@sanitize@url [0]{\catcode `\\12\catcode `\$12\catcode
		`\&12\catcode `\#12\catcode `\^12\catcode `\_12\catcode `\%12\relax}%
	\providecommand \@@startlink[1]{}%
	\providecommand \@@endlink[0]{}%
	\providecommand \url  [0]{\begingroup\@sanitize@url \@url }%
	\providecommand \@url [1]{\endgroup\@href {#1}{\urlprefix }}%
	\providecommand \urlprefix  [0]{URL }%
	\providecommand \Eprint [0]{\href }%
	\providecommand \doibase [0]{https://doi.org/}%
	\providecommand \selectlanguage [0]{\@gobble}%
	\providecommand \bibinfo  [0]{\@secondoftwo}%
	\providecommand \bibfield  [0]{\@secondoftwo}%
	\providecommand \translation [1]{[#1]}%
	\providecommand \BibitemOpen [0]{}%
	\providecommand \bibitemStop [0]{}%
	\providecommand \bibitemNoStop [0]{.\EOS\space}%
	\providecommand \EOS [0]{\spacefactor3000\relax}%
	\providecommand \BibitemShut  [1]{\csname bibitem#1\endcsname}%
	\let\auto@bib@innerbib\@empty
	\bibitem [{\citenamefont {Back}\ \emph {et~al.}(1998)\citenamefont {Back},
		\citenamefont {Weller}, \citenamefont {Heidmann}, \citenamefont {Mauri},
		\citenamefont {Guarisco}, \citenamefont {Garwin},\ and\ \citenamefont
		{Siegmann}}]{Back.1998}%
	\BibitemOpen
	\bibfield  {author} {\bibinfo {author} {\bibfnamefont {C.~H.}\ \bibnamefont
			{Back}}, \bibinfo {author} {\bibfnamefont {D.}~\bibnamefont {Weller}},
		\bibinfo {author} {\bibfnamefont {J.}~\bibnamefont {Heidmann}}, \bibinfo
		{author} {\bibfnamefont {D.}~\bibnamefont {Mauri}}, \bibinfo {author}
		{\bibfnamefont {D.}~\bibnamefont {Guarisco}}, \bibinfo {author}
		{\bibfnamefont {E.~L.}\ \bibnamefont {Garwin}},\ and\ \bibinfo {author}
		{\bibfnamefont {H.~C.}\ \bibnamefont {Siegmann}},\ }\bibfield  {title} {\emph
		{\bibinfo {title} {Magnetization {R}eversal in {U}ltrashort {M}agnetic
				{F}ield {P}ulses}},\ }\href {https://doi.org/10.1103/PhysRevLett.81.3251}
	{\bibfield  {journal} {\bibinfo  {journal} {Phys. Rev. Lett.}\ }\textbf
		{\bibinfo {volume} {81}},\ \bibinfo {pages} {3251} (\bibinfo {year}
		{1998})}\BibitemShut {NoStop}%
	\bibitem [{\citenamefont {Schumacher}\ \emph {et~al.}(2003)\citenamefont
		{Schumacher}, \citenamefont {Chappert}, \citenamefont {Sousa}, \citenamefont
		{Freitas},\ and\ \citenamefont {Miltat}}]{Schumacher.2003}%
	\BibitemOpen
	\bibfield  {author} {\bibinfo {author} {\bibfnamefont {H.~W.}\ \bibnamefont
			{Schumacher}}, \bibinfo {author} {\bibfnamefont {C.}~\bibnamefont
			{Chappert}}, \bibinfo {author} {\bibfnamefont {R.~C.}\ \bibnamefont {Sousa}},
		\bibinfo {author} {\bibfnamefont {P.~P.}\ \bibnamefont {Freitas}},\ and\
		\bibinfo {author} {\bibfnamefont {J.}~\bibnamefont {Miltat}},\ }\bibfield
	{title} {\emph {\bibinfo {title} {Quasiballistic {M}agnetization
				{R}eversal}},\ }\href {https://doi.org/10.1103/PhysRevLett.90.017204}
	{\bibfield  {journal} {\bibinfo  {journal} {Phys. Rev. Lett.}\ }\textbf
		{\bibinfo {volume} {90}},\ \bibinfo {pages} {017204} (\bibinfo {year}
		{2003})}\BibitemShut {NoStop}%
	\bibitem [{\citenamefont {Tudosa}\ \emph {et~al.}(2004)\citenamefont {Tudosa},
		\citenamefont {Stamm}, \citenamefont {Kashuba}, \citenamefont {King},
		\citenamefont {Siegmann}, \citenamefont {St{\"o}hr}, \citenamefont {Ju},
		\citenamefont {Lu},\ and\ \citenamefont {Weller}}]{Tudosa.2004}%
	\BibitemOpen
	\bibfield  {author} {\bibinfo {author} {\bibfnamefont {I.}~\bibnamefont
			{Tudosa}}, \bibinfo {author} {\bibfnamefont {C.}~\bibnamefont {Stamm}},
		\bibinfo {author} {\bibfnamefont {A.~B.}\ \bibnamefont {Kashuba}}, \bibinfo
		{author} {\bibfnamefont {F.}~\bibnamefont {King}}, \bibinfo {author}
		{\bibfnamefont {H.~C.}\ \bibnamefont {Siegmann}}, \bibinfo {author}
		{\bibfnamefont {J.}~\bibnamefont {St{\"o}hr}}, \bibinfo {author}
		{\bibfnamefont {G.}~\bibnamefont {Ju}}, \bibinfo {author} {\bibfnamefont
			{B.}~\bibnamefont {Lu}},\ and\ \bibinfo {author} {\bibfnamefont
			{D.}~\bibnamefont {Weller}},\ }\bibfield  {title} {\emph {\bibinfo {title}
			{The ultimate speed of magnetic switching in granular recording media}},\
	}\href {https://doi.org/10.1038/nature02438} {\bibfield  {journal} {\bibinfo
			{journal} {Nature}\ }\textbf {\bibinfo {volume} {428}},\ \bibinfo {pages}
		{831} (\bibinfo {year} {2004})}\BibitemShut {NoStop}%
	\bibitem [{\citenamefont {Beaurepaire}\ \emph {et~al.}(1996)\citenamefont
		{Beaurepaire}, \citenamefont {Merle}, \citenamefont {Daunois},\ and\
		\citenamefont {Bigot}}]{Beaurepaire.1996}%
	\BibitemOpen
	\bibfield  {author} {\bibinfo {author} {\bibfnamefont {E.}~\bibnamefont
			{Beaurepaire}}, \bibinfo {author} {\bibfnamefont {J.-C.}\ \bibnamefont
			{Merle}}, \bibinfo {author} {\bibfnamefont {A.}~\bibnamefont {Daunois}},\
		and\ \bibinfo {author} {\bibfnamefont {J.-Y.}\ \bibnamefont {Bigot}},\
	}\bibfield  {title} {\emph {\bibinfo {title} {Ultrafast spin dynamics in
				ferromagnetic nickel}},\ }\href {https://doi.org/10.1103/PhysRevLett.76.4250}
	{\bibfield  {journal} {\bibinfo  {journal} {Phys. Rev. Lett.}\ }\textbf
		{\bibinfo {volume} {76}},\ \bibinfo {pages} {4250} (\bibinfo {year}
		{1996})}\BibitemShut {NoStop}%
	\bibitem [{\citenamefont {Koopmans}\ \emph {et~al.}(2005)\citenamefont
		{Koopmans}, \citenamefont {Ruigrok}, \citenamefont {Longa},\ and\
		\citenamefont {de~Jonge}}]{Koopmans.2005}%
	\BibitemOpen
	\bibfield  {author} {\bibinfo {author} {\bibfnamefont {B.}~\bibnamefont
			{Koopmans}}, \bibinfo {author} {\bibfnamefont {J.~J.~M.}\ \bibnamefont
			{Ruigrok}}, \bibinfo {author} {\bibfnamefont {F.~D.}\ \bibnamefont {Longa}},\
		and\ \bibinfo {author} {\bibfnamefont {W.~J.~M.}\ \bibnamefont {de~Jonge}},\
	}\bibfield  {title} {\emph {\bibinfo {title} {Unifying {U}ltrafast
				{M}agnetization {D}ynamics}},\ }\href
	{https://doi.org/10.1103/PhysRevLett.95.267207} {\bibfield  {journal}
		{\bibinfo  {journal} {Phys. Rev. Lett.}\ }\textbf {\bibinfo {volume} {95}},\
		\bibinfo {pages} {267207} (\bibinfo {year} {2005})}\BibitemShut {NoStop}%
	\bibitem [{\citenamefont {Cinchetti}\ \emph {et~al.}(2006)\citenamefont
		{Cinchetti}, \citenamefont {{S{\'a}nchez Albaneda}}, \citenamefont
		{Hoffmann}, \citenamefont {Roth}, \citenamefont {W{\"u}stenberg},
		\citenamefont {Krauss}, \citenamefont {Andreyev}, \citenamefont {Schneider},
		\citenamefont {Bauer},\ and\ \citenamefont {Aeschlimann}}]{Cinchetti.2006}%
	\BibitemOpen
	\bibfield  {author} {\bibinfo {author} {\bibfnamefont {M.}~\bibnamefont
			{Cinchetti}}, \bibinfo {author} {\bibfnamefont {M.}~\bibnamefont
			{{S{\'a}nchez Albaneda}}}, \bibinfo {author} {\bibfnamefont {D.}~\bibnamefont
			{Hoffmann}}, \bibinfo {author} {\bibfnamefont {T.}~\bibnamefont {Roth}},
		\bibinfo {author} {\bibfnamefont {J.-P.}\ \bibnamefont {W{\"u}stenberg}},
		\bibinfo {author} {\bibfnamefont {M.}~\bibnamefont {Krauss}}, \bibinfo
		{author} {\bibfnamefont {O.}~\bibnamefont {Andreyev}}, \bibinfo {author}
		{\bibfnamefont {H.~C.}\ \bibnamefont {Schneider}}, \bibinfo {author}
		{\bibfnamefont {M.}~\bibnamefont {Bauer}},\ and\ \bibinfo {author}
		{\bibfnamefont {M.}~\bibnamefont {Aeschlimann}},\ }\bibfield  {title} {\emph
		{\bibinfo {title} {Spin-flip processes and ultrafast magnetization dynamics
				in {C}o: {U}nifying the microscopic and macroscopic view of femtosecond
				magnetism}},\ }\href {https://doi.org/10.1103/PhysRevLett.97.177201}
	{\bibfield  {journal} {\bibinfo  {journal} {Phys. Rev. Lett.}\ }\textbf
		{\bibinfo {volume} {97}},\ \bibinfo {pages} {177201} (\bibinfo {year}
		{2006})}\BibitemShut {NoStop}%
	\bibitem [{\citenamefont {Bigot}\ \emph {et~al.}(2009)\citenamefont {Bigot},
		\citenamefont {Vomir},\ and\ \citenamefont {Beaurepaire}}]{Bigot.2009}%
	\BibitemOpen
	\bibfield  {author} {\bibinfo {author} {\bibfnamefont {J.-Y.}\ \bibnamefont
			{Bigot}}, \bibinfo {author} {\bibfnamefont {M.}~\bibnamefont {Vomir}},\ and\
		\bibinfo {author} {\bibfnamefont {E.}~\bibnamefont {Beaurepaire}},\
	}\bibfield  {title} {\emph {\bibinfo {title} {Coherent ultrafast magnetism
				induced by femtosecond laser pulses}},\ }\href
	{https://doi.org/10.1038/nphys1285} {\bibfield  {journal} {\bibinfo
			{journal} {Nat. Phys.}\ }\textbf {\bibinfo {volume} {5}},\ \bibinfo {pages}
		{515} (\bibinfo {year} {2009})}\BibitemShut {NoStop}%
	\bibitem [{\citenamefont {Koopmans}\ \emph {et~al.}(2010)\citenamefont
		{Koopmans}, \citenamefont {Malinowski}, \citenamefont {{Dalla Longa}},
		\citenamefont {Steiauf}, \citenamefont {F{\"a}hnle}, \citenamefont {Roth},
		\citenamefont {Cinchetti},\ and\ \citenamefont
		{Aeschlimann}}]{Koopmans.2010}%
	\BibitemOpen
	\bibfield  {author} {\bibinfo {author} {\bibfnamefont {B.}~\bibnamefont
			{Koopmans}}, \bibinfo {author} {\bibfnamefont {G.}~\bibnamefont
			{Malinowski}}, \bibinfo {author} {\bibfnamefont {F.}~\bibnamefont {{Dalla
					Longa}}}, \bibinfo {author} {\bibfnamefont {D.}~\bibnamefont {Steiauf}},
		\bibinfo {author} {\bibfnamefont {M.}~\bibnamefont {F{\"a}hnle}}, \bibinfo
		{author} {\bibfnamefont {T.}~\bibnamefont {Roth}}, \bibinfo {author}
		{\bibfnamefont {M.}~\bibnamefont {Cinchetti}},\ and\ \bibinfo {author}
		{\bibfnamefont {M.}~\bibnamefont {Aeschlimann}},\ }\bibfield  {title} {\emph
		{\bibinfo {title} {Explaining the paradoxical diversity of ultrafast
				laser-induced demagnetization}},\ }\href {https://doi.org/10.1038/nmat2593}
	{\bibfield  {journal} {\bibinfo  {journal} {Nat. Mater.}\ }\textbf {\bibinfo
			{volume} {9}},\ \bibinfo {pages} {259} (\bibinfo {year} {2010})}\BibitemShut
	{NoStop}%
	\bibitem [{\citenamefont {Kirilyuk}\ \emph {et~al.}(2010)\citenamefont
		{Kirilyuk}, \citenamefont {Kimel},\ and\ \citenamefont
		{Rasing}}]{Kirilyuk.2010}%
	\BibitemOpen
	\bibfield  {author} {\bibinfo {author} {\bibfnamefont {A.}~\bibnamefont
			{Kirilyuk}}, \bibinfo {author} {\bibfnamefont {A.~V.}\ \bibnamefont
			{Kimel}},\ and\ \bibinfo {author} {\bibfnamefont {T.}~\bibnamefont
			{Rasing}},\ }\bibfield  {title} {\emph {\bibinfo {title} {Ultrafast optical
				manipulation of magnetic order}},\ }\href
	{https://doi.org/10.1103/RevModPhys.82.2731} {\bibfield  {journal} {\bibinfo
			{journal} {Rev. Mod. Phys.}\ }\textbf {\bibinfo {volume} {82}},\ \bibinfo
		{pages} {2731} (\bibinfo {year} {2010})}\BibitemShut {NoStop}%
	\bibitem [{\citenamefont {Lambert}\ \emph {et~al.}(2014)\citenamefont
		{Lambert}, \citenamefont {Mangin}, \citenamefont {Varaprasad}, \citenamefont
		{Takahashi}, \citenamefont {Hehn}, \citenamefont {Cinchetti}, \citenamefont
		{Malinowski}, \citenamefont {Hono}, \citenamefont {Fainman}, \citenamefont
		{Aeschlimann},\ and\ \citenamefont {Fullerton}}]{Lambert.2014}%
	\BibitemOpen
	\bibfield  {author} {\bibinfo {author} {\bibfnamefont {C.-H.}\ \bibnamefont
			{Lambert}}, \bibinfo {author} {\bibfnamefont {S.}~\bibnamefont {Mangin}},
		\bibinfo {author} {\bibfnamefont {B.~S. D. C.~S.}\ \bibnamefont
			{Varaprasad}}, \bibinfo {author} {\bibfnamefont {Y.~K.}\ \bibnamefont
			{Takahashi}}, \bibinfo {author} {\bibfnamefont {M.}~\bibnamefont {Hehn}},
		\bibinfo {author} {\bibfnamefont {M.}~\bibnamefont {Cinchetti}}, \bibinfo
		{author} {\bibfnamefont {G.}~\bibnamefont {Malinowski}}, \bibinfo {author}
		{\bibfnamefont {K.}~\bibnamefont {Hono}}, \bibinfo {author} {\bibfnamefont
			{Y.}~\bibnamefont {Fainman}}, \bibinfo {author} {\bibfnamefont
			{M.}~\bibnamefont {Aeschlimann}},\ and\ \bibinfo {author} {\bibfnamefont
			{E.~E.}\ \bibnamefont {Fullerton}},\ }\bibfield  {title} {\emph {\bibinfo
			{title} {All-optical control of ferromagnetic thin films and
				nanostructures}},\ }\href {https://doi.org/10.1126/science.1253493}
	{\bibfield  {journal} {\bibinfo  {journal} {Science}\ }\textbf {\bibinfo
			{volume} {345}},\ \bibinfo {pages} {1337} (\bibinfo {year}
		{2014})}\BibitemShut {NoStop}%
	\bibitem [{\citenamefont {Walowski}\ and\ \citenamefont
		{M\"unzenberg}(2016)}]{Walowski.2016}%
	\BibitemOpen
	\bibfield  {author} {\bibinfo {author} {\bibfnamefont {J.}~\bibnamefont
			{Walowski}}\ and\ \bibinfo {author} {\bibfnamefont {M.}~\bibnamefont
			{M\"unzenberg}},\ }\bibfield  {title} {\emph {\bibinfo {title} {Perspective:
				{U}ltrafast magnetism and {TH}z spintronics}},\ }\href
	{https://doi.org/10.1063/1.4958846} {\bibfield  {journal} {\bibinfo
			{journal} {J. Appl. Phys.}\ }\textbf {\bibinfo {volume} {120}},\ \bibinfo
		{pages} {140901} (\bibinfo {year} {2016})}\BibitemShut {NoStop}%
	\bibitem [{\citenamefont {Stamm}\ \emph {et~al.}(2007)\citenamefont {Stamm},
		\citenamefont {Kachel}, \citenamefont {Pontius}, \citenamefont {Mitzner},
		\citenamefont {Quast}, \citenamefont {Holldack}, \citenamefont {Khan},
		\citenamefont {Lupulescu}, \citenamefont {Aziz}, \citenamefont {Wietstruk},
		\citenamefont {D{\"u}rr},\ and\ \citenamefont {Eberhardt}}]{Stamm.2007}%
	\BibitemOpen
	\bibfield  {author} {\bibinfo {author} {\bibfnamefont {C.}~\bibnamefont
			{Stamm}}, \bibinfo {author} {\bibfnamefont {T.}~\bibnamefont {Kachel}},
		\bibinfo {author} {\bibfnamefont {N.}~\bibnamefont {Pontius}}, \bibinfo
		{author} {\bibfnamefont {R.}~\bibnamefont {Mitzner}}, \bibinfo {author}
		{\bibfnamefont {T.}~\bibnamefont {Quast}}, \bibinfo {author} {\bibfnamefont
			{K.}~\bibnamefont {Holldack}}, \bibinfo {author} {\bibfnamefont
			{S.}~\bibnamefont {Khan}}, \bibinfo {author} {\bibfnamefont {C.}~\bibnamefont
			{Lupulescu}}, \bibinfo {author} {\bibfnamefont {E.~F.}\ \bibnamefont {Aziz}},
		\bibinfo {author} {\bibfnamefont {M.}~\bibnamefont {Wietstruk}}, \bibinfo
		{author} {\bibfnamefont {H.~A.}\ \bibnamefont {D{\"u}rr}},\ and\ \bibinfo
		{author} {\bibfnamefont {W.}~\bibnamefont {Eberhardt}},\ }\bibfield  {title}
	{\emph {\bibinfo {title} {Femtosecond modification of electron localization
				and transfer of angular momentum in nickel}},\ }\href
	{https://doi.org/10.1038/nmat1985} {\bibfield  {journal} {\bibinfo  {journal}
			{Nat. Mater.}\ }\textbf {\bibinfo {volume} {6}},\ \bibinfo {pages} {740}
		(\bibinfo {year} {2007})}\BibitemShut {NoStop}%
	\bibitem [{\citenamefont {Krau\ss{}}\ \emph {et~al.}(2009)\citenamefont
		{Krau\ss{}}, \citenamefont {Roth}, \citenamefont {Alebrand}, \citenamefont
		{Steil}, \citenamefont {Cinchetti}, \citenamefont {Aeschlimann},\ and\
		\citenamefont {Schneider}}]{Krau.2009}%
	\BibitemOpen
	\bibfield  {author} {\bibinfo {author} {\bibfnamefont {M.}~\bibnamefont
			{Krau\ss{}}}, \bibinfo {author} {\bibfnamefont {T.}~\bibnamefont {Roth}},
		\bibinfo {author} {\bibfnamefont {S.}~\bibnamefont {Alebrand}}, \bibinfo
		{author} {\bibfnamefont {D.}~\bibnamefont {Steil}}, \bibinfo {author}
		{\bibfnamefont {M.}~\bibnamefont {Cinchetti}}, \bibinfo {author}
		{\bibfnamefont {M.}~\bibnamefont {Aeschlimann}},\ and\ \bibinfo {author}
		{\bibfnamefont {H.~C.}\ \bibnamefont {Schneider}},\ }\bibfield  {title}
	{\emph {\bibinfo {title} {Ultrafast demagnetization of ferromagnetic
				transition metals: The role of the {C}oulomb interaction}},\ }\href
	{https://doi.org/10.1103/PhysRevB.80.180407} {\bibfield  {journal} {\bibinfo
			{journal} {Phys. Rev. B}\ }\textbf {\bibinfo {volume} {80}},\ \bibinfo
		{pages} {180407} (\bibinfo {year} {2009})}\BibitemShut {NoStop}%
	\bibitem [{\citenamefont {Turgut}\ \emph {et~al.}(2016)\citenamefont {Turgut},
		\citenamefont {Zusin}, \citenamefont {Legut}, \citenamefont {Carva},
		\citenamefont {Knut}, \citenamefont {Shaw}, \citenamefont {Chen},
		\citenamefont {Tao}, \citenamefont {Nembach}, \citenamefont {Silva},
		\citenamefont {Mathias}, \citenamefont {Aeschlimann}, \citenamefont
		{Oppeneer}, \citenamefont {Kapteyn}, \citenamefont {Murnane},\ and\
		\citenamefont {Grychtol}}]{Turgut.2016}%
	\BibitemOpen
	\bibfield  {author} {\bibinfo {author} {\bibfnamefont {E.}~\bibnamefont
			{Turgut}}, \bibinfo {author} {\bibfnamefont {D.}~\bibnamefont {Zusin}},
		\bibinfo {author} {\bibfnamefont {D.}~\bibnamefont {Legut}}, \bibinfo
		{author} {\bibfnamefont {K.}~\bibnamefont {Carva}}, \bibinfo {author}
		{\bibfnamefont {R.}~\bibnamefont {Knut}}, \bibinfo {author} {\bibfnamefont
			{J.~M.}\ \bibnamefont {Shaw}}, \bibinfo {author} {\bibfnamefont
			{C.}~\bibnamefont {Chen}}, \bibinfo {author} {\bibfnamefont {Z.}~\bibnamefont
			{Tao}}, \bibinfo {author} {\bibfnamefont {H.~T.}\ \bibnamefont {Nembach}},
		\bibinfo {author} {\bibfnamefont {T.~J.}\ \bibnamefont {Silva}}, \bibinfo
		{author} {\bibfnamefont {S.}~\bibnamefont {Mathias}}, \bibinfo {author}
		{\bibfnamefont {M.}~\bibnamefont {Aeschlimann}}, \bibinfo {author}
		{\bibfnamefont {P.~M.}\ \bibnamefont {Oppeneer}}, \bibinfo {author}
		{\bibfnamefont {H.~C.}\ \bibnamefont {Kapteyn}}, \bibinfo {author}
		{\bibfnamefont {M.~M.}\ \bibnamefont {Murnane}},\ and\ \bibinfo {author}
		{\bibfnamefont {P.}~\bibnamefont {Grychtol}},\ }\bibfield  {title} {\emph
		{\bibinfo {title} {Stoner versus {H}eisenberg: {U}ltrafast exchange reduction
				and magnon generation during laser-induced demagnetization}},\ }\href
	{https://doi.org/10.1103/PhysRevB.94.220408} {\bibfield  {journal} {\bibinfo
			{journal} {Phys. Rev. B}\ }\textbf {\bibinfo {volume} {94}},\ \bibinfo
		{pages} {220408} (\bibinfo {year} {2016})}\BibitemShut {NoStop}%
	\bibitem [{\citenamefont {Battiato}\ \emph {et~al.}(2010)\citenamefont
		{Battiato}, \citenamefont {Carva},\ and\ \citenamefont
		{Oppeneer}}]{Battiato.2010}%
	\BibitemOpen
	\bibfield  {author} {\bibinfo {author} {\bibfnamefont {M.}~\bibnamefont
			{Battiato}}, \bibinfo {author} {\bibfnamefont {K.}~\bibnamefont {Carva}},\
		and\ \bibinfo {author} {\bibfnamefont {P.~M.}\ \bibnamefont {Oppeneer}},\
	}\bibfield  {title} {\emph {\bibinfo {title} {Superdiffusive spin transport
				as a mechanism of ultrafast demagnetization}},\ }\href
	{https://doi.org/10.1103/PhysRevLett.105.027203} {\bibfield  {journal}
		{\bibinfo  {journal} {Phys. Rev. Lett.}\ }\textbf {\bibinfo {volume} {105}},\
		\bibinfo {pages} {027203} (\bibinfo {year} {2010})}\BibitemShut {NoStop}%
	\bibitem [{\citenamefont {Eschenlohr}\ \emph {et~al.}(2013)\citenamefont
		{Eschenlohr}, \citenamefont {Battiato}, \citenamefont {Maldonado},
		\citenamefont {Pontius}, \citenamefont {Kachel}, \citenamefont {Holldack},
		\citenamefont {Mitzner}, \citenamefont {F{\"o}hlisch}, \citenamefont
		{Oppeneer},\ and\ \citenamefont {Stamm}}]{Eschenlohr.2013}%
	\BibitemOpen
	\bibfield  {author} {\bibinfo {author} {\bibfnamefont {A.}~\bibnamefont
			{Eschenlohr}}, \bibinfo {author} {\bibfnamefont {M.}~\bibnamefont
			{Battiato}}, \bibinfo {author} {\bibfnamefont {P.}~\bibnamefont {Maldonado}},
		\bibinfo {author} {\bibfnamefont {N.}~\bibnamefont {Pontius}}, \bibinfo
		{author} {\bibfnamefont {T.}~\bibnamefont {Kachel}}, \bibinfo {author}
		{\bibfnamefont {K.}~\bibnamefont {Holldack}}, \bibinfo {author}
		{\bibfnamefont {R.}~\bibnamefont {Mitzner}}, \bibinfo {author} {\bibfnamefont
			{A.}~\bibnamefont {F{\"o}hlisch}}, \bibinfo {author} {\bibfnamefont {P.~M.}\
			\bibnamefont {Oppeneer}},\ and\ \bibinfo {author} {\bibfnamefont
			{C.}~\bibnamefont {Stamm}},\ }\bibfield  {title} {\emph {\bibinfo {title}
			{Ultrafast spin transport as key to femtosecond demagnetization}},\ }\href
	{https://doi.org/10.1038/nmat3546} {\bibfield  {journal} {\bibinfo  {journal}
			{Nat. Mater.}\ }\textbf {\bibinfo {volume} {12}},\ \bibinfo {pages} {332}
		(\bibinfo {year} {2013})}\BibitemShut {NoStop}%
	\bibitem [{\citenamefont {Rudolf}\ \emph {et~al.}(2012)\citenamefont {Rudolf},
		\citenamefont {La-o Vorakiat}, \citenamefont {Battiato}, \citenamefont
		{Adam}, \citenamefont {Shaw}, \citenamefont {Turgut}, \citenamefont
		{Maldonado}, \citenamefont {Mathias}, \citenamefont {Grychtol}, \citenamefont
		{Nembach}, \citenamefont {Silva}, \citenamefont {Aeschlimann}, \citenamefont
		{Kapteyn}, \citenamefont {Murnane}, \citenamefont {Schneider},\ and\
		\citenamefont {Oppeneer}}]{Rudolf.2012}%
	\BibitemOpen
	\bibfield  {author} {\bibinfo {author} {\bibfnamefont {D.}~\bibnamefont
			{Rudolf}}, \bibinfo {author} {\bibfnamefont {C.}~\bibnamefont {La-o
				Vorakiat}}, \bibinfo {author} {\bibfnamefont {M.}~\bibnamefont {Battiato}},
		\bibinfo {author} {\bibfnamefont {R.}~\bibnamefont {Adam}}, \bibinfo {author}
		{\bibfnamefont {J.~M.}\ \bibnamefont {Shaw}}, \bibinfo {author}
		{\bibfnamefont {E.}~\bibnamefont {Turgut}}, \bibinfo {author} {\bibfnamefont
			{P.}~\bibnamefont {Maldonado}}, \bibinfo {author} {\bibfnamefont
			{S.}~\bibnamefont {Mathias}}, \bibinfo {author} {\bibfnamefont
			{P.}~\bibnamefont {Grychtol}}, \bibinfo {author} {\bibfnamefont {H.~T.}\
			\bibnamefont {Nembach}}, \bibinfo {author} {\bibfnamefont {T.~J.}\
			\bibnamefont {Silva}}, \bibinfo {author} {\bibfnamefont {M.}~\bibnamefont
			{Aeschlimann}}, \bibinfo {author} {\bibfnamefont {H.~C.}\ \bibnamefont
			{Kapteyn}}, \bibinfo {author} {\bibfnamefont {M.~M.}\ \bibnamefont
			{Murnane}}, \bibinfo {author} {\bibfnamefont {C.~M.}\ \bibnamefont
			{Schneider}},\ and\ \bibinfo {author} {\bibfnamefont {P.~M.}\ \bibnamefont
			{Oppeneer}},\ }\bibfield  {title} {\emph {\bibinfo {title} {Ultrafast
				magnetization enhancement in metallic multilayers driven by superdiffusive
				spin current}},\ }\href {https://doi.org/10.1038/ncomms2029} {\bibfield
		{journal} {\bibinfo  {journal} {Nat. Commun.}\ }\textbf {\bibinfo {volume}
			{3}},\ \bibinfo {pages} {1037} (\bibinfo {year} {2012})}\BibitemShut
	{NoStop}%
	\bibitem [{\citenamefont {Turgut}\ \emph {et~al.}(2013)\citenamefont {Turgut},
		\citenamefont {La-o vorakiat}, \citenamefont {Shaw}, \citenamefont
		{Grychtol}, \citenamefont {Nembach}, \citenamefont {Rudolf}, \citenamefont
		{Adam}, \citenamefont {Aeschlimann}, \citenamefont {Schneider}, \citenamefont
		{Silva}, \citenamefont {Murnane}, \citenamefont {Kapteyn},\ and\
		\citenamefont {Mathias}}]{Turgut.2013}%
	\BibitemOpen
	\bibfield  {author} {\bibinfo {author} {\bibfnamefont {E.}~\bibnamefont
			{Turgut}}, \bibinfo {author} {\bibfnamefont {C.}~\bibnamefont {La-o
				vorakiat}}, \bibinfo {author} {\bibfnamefont {J.~M.}\ \bibnamefont {Shaw}},
		\bibinfo {author} {\bibfnamefont {P.}~\bibnamefont {Grychtol}}, \bibinfo
		{author} {\bibfnamefont {H.~T.}\ \bibnamefont {Nembach}}, \bibinfo {author}
		{\bibfnamefont {D.}~\bibnamefont {Rudolf}}, \bibinfo {author} {\bibfnamefont
			{R.}~\bibnamefont {Adam}}, \bibinfo {author} {\bibfnamefont {M.}~\bibnamefont
			{Aeschlimann}}, \bibinfo {author} {\bibfnamefont {C.~M.}\ \bibnamefont
			{Schneider}}, \bibinfo {author} {\bibfnamefont {T.~J.}\ \bibnamefont
			{Silva}}, \bibinfo {author} {\bibfnamefont {M.~M.}\ \bibnamefont {Murnane}},
		\bibinfo {author} {\bibfnamefont {H.~C.}\ \bibnamefont {Kapteyn}},\ and\
		\bibinfo {author} {\bibfnamefont {S.}~\bibnamefont {Mathias}},\ }\bibfield
	{title} {\emph {\bibinfo {title} {Controlling the {C}ompetition between
				{O}ptically {I}nduced {U}ltrafast {S}pin-{F}lip {S}cattering and {S}pin
				{T}ransport in {M}agnetic {M}ultilayers}},\ }\href
	{https://doi.org/10.1103/PhysRevLett.110.197201} {\bibfield  {journal}
		{\bibinfo  {journal} {Phys. Rev. Lett.}\ }\textbf {\bibinfo {volume} {110}},\
		\bibinfo {pages} {197201} (\bibinfo {year} {2013})}\BibitemShut {NoStop}%
	\bibitem [{\citenamefont {Meiklejohn}\ and\ \citenamefont
		{Bean}(1956)}]{Meiklejohn.1956}%
	\BibitemOpen
	\bibfield  {author} {\bibinfo {author} {\bibfnamefont {W.~H.}\ \bibnamefont
			{Meiklejohn}}\ and\ \bibinfo {author} {\bibfnamefont {C.~P.}\ \bibnamefont
			{Bean}},\ }\bibfield  {title} {\emph {\bibinfo {title} {New {M}agnetic
				{A}nisotropy}},\ }\href {https://doi.org/10.1103/PhysRev.102.1413} {\bibfield
		{journal} {\bibinfo  {journal} {Phys. Rev.}\ }\textbf {\bibinfo {volume}
			{102}},\ \bibinfo {pages} {1413} (\bibinfo {year} {1956})}\BibitemShut
	{NoStop}%
	\bibitem [{\citenamefont {Nogu{\'e}s}\ and\ \citenamefont
		{Schuller}(1999)}]{Nogues.1999}%
	\BibitemOpen
	\bibfield  {author} {\bibinfo {author} {\bibfnamefont {J.}~\bibnamefont
			{Nogu{\'e}s}}\ and\ \bibinfo {author} {\bibfnamefont {I.~K.}\ \bibnamefont
			{Schuller}},\ }\bibfield  {title} {\emph {\bibinfo {title} {Exchange bias}},\
	}\href {https://doi.org/10.1016/S0304-8853(98)00266-2} {\bibfield  {journal}
		{\bibinfo  {journal} {J. Magn. Magn. Mater.}\ }\textbf {\bibinfo {volume}
			{192}},\ \bibinfo {pages} {203} (\bibinfo {year} {1999})}\BibitemShut
	{NoStop}%
	\bibitem [{\citenamefont {Ju}\ \emph {et~al.}(1998)\citenamefont {Ju},
		\citenamefont {Nurmikko}, \citenamefont {Farrow}, \citenamefont {Marks},
		\citenamefont {Carey},\ and\ \citenamefont {Gurney}}]{Ju.1998}%
	\BibitemOpen
	\bibfield  {author} {\bibinfo {author} {\bibfnamefont {G.}~\bibnamefont
			{Ju}}, \bibinfo {author} {\bibfnamefont {A.~V.}\ \bibnamefont {Nurmikko}},
		\bibinfo {author} {\bibfnamefont {R.~F.~C.}\ \bibnamefont {Farrow}}, \bibinfo
		{author} {\bibfnamefont {R.~F.}\ \bibnamefont {Marks}}, \bibinfo {author}
		{\bibfnamefont {M.~J.}\ \bibnamefont {Carey}},\ and\ \bibinfo {author}
		{\bibfnamefont {B.~A.}\ \bibnamefont {Gurney}},\ }\bibfield  {title} {\emph
		{\bibinfo {title} {Ultrafast optical modulation of an exchange biased
				ferromagnetic/antiferromagnetic bilayer}},\ }\href
	{https://doi.org/10.1103/PhysRevB.58.11857} {\bibfield  {journal} {\bibinfo
			{journal} {Phys. Rev. B}\ }\textbf {\bibinfo {volume} {58}},\ \bibinfo
		{pages} {11857} (\bibinfo {year} {1998})}\BibitemShut {NoStop}%
	\bibitem [{\citenamefont {Ju}\ \emph {et~al.}(1999)\citenamefont {Ju},
		\citenamefont {Nurmikko}, \citenamefont {Farrow}, \citenamefont {Marks},
		\citenamefont {Carey},\ and\ \citenamefont {Gurney}}]{Ju.1999}%
	\BibitemOpen
	\bibfield  {author} {\bibinfo {author} {\bibfnamefont {G.}~\bibnamefont
			{Ju}}, \bibinfo {author} {\bibfnamefont {A.~V.}\ \bibnamefont {Nurmikko}},
		\bibinfo {author} {\bibfnamefont {R.~F.~C.}\ \bibnamefont {Farrow}}, \bibinfo
		{author} {\bibfnamefont {R.~F.}\ \bibnamefont {Marks}}, \bibinfo {author}
		{\bibfnamefont {M.~J.}\ \bibnamefont {Carey}},\ and\ \bibinfo {author}
		{\bibfnamefont {B.~A.}\ \bibnamefont {Gurney}},\ }\bibfield  {title} {\emph
		{\bibinfo {title} {Ultrafast {T}ime {R}esolved {P}hotoinduced {M}agnetization
				{R}otation in a {F}erromagnetic/{A}ntiferromagnetic {E}xchange {C}oupled
				{S}ystem}},\ }\href {https://doi.org/10.1103/PhysRevLett.82.3705} {\bibfield
		{journal} {\bibinfo  {journal} {Phys. Rev. Lett.}\ }\textbf {\bibinfo
			{volume} {82}},\ \bibinfo {pages} {3705} (\bibinfo {year}
		{1999})}\BibitemShut {NoStop}%
	\bibitem [{\citenamefont {Ju}\ \emph {et~al.}(2000)\citenamefont {Ju},
		\citenamefont {Chen}, \citenamefont {Nurmikko}, \citenamefont {Farrow},
		\citenamefont {Marks}, \citenamefont {Carey},\ and\ \citenamefont
		{Gurney}}]{Ju.2000}%
	\BibitemOpen
	\bibfield  {author} {\bibinfo {author} {\bibfnamefont {G.}~\bibnamefont
			{Ju}}, \bibinfo {author} {\bibfnamefont {L.}~\bibnamefont {Chen}}, \bibinfo
		{author} {\bibfnamefont {A.~V.}\ \bibnamefont {Nurmikko}}, \bibinfo {author}
		{\bibfnamefont {R.~F.~C.}\ \bibnamefont {Farrow}}, \bibinfo {author}
		{\bibfnamefont {R.~F.}\ \bibnamefont {Marks}}, \bibinfo {author}
		{\bibfnamefont {M.~J.}\ \bibnamefont {Carey}},\ and\ \bibinfo {author}
		{\bibfnamefont {B.~A.}\ \bibnamefont {Gurney}},\ }\bibfield  {title} {\emph
		{\bibinfo {title} {Coherent magnetization rotation induced by optical
				modulation in ferromagnetic/antiferromagnetic exchange-coupled bilayers}},\
	}\href {https://doi.org/10.1103/PhysRevB.62.1171} {\bibfield  {journal}
		{\bibinfo  {journal} {Phys. Rev. B}\ }\textbf {\bibinfo {volume} {62}},\
		\bibinfo {pages} {1171} (\bibinfo {year} {2000})}\BibitemShut {NoStop}%
	\bibitem [{\citenamefont {Dalla~Longa}\ \emph {et~al.}(2010)\citenamefont
		{Dalla~Longa}, \citenamefont {Kohlhepp}, \citenamefont {de~Jonge},\ and\
		\citenamefont {Koopmans}}]{DallaLonga.2010}%
	\BibitemOpen
	\bibfield  {author} {\bibinfo {author} {\bibfnamefont {F.}~\bibnamefont
			{Dalla~Longa}}, \bibinfo {author} {\bibfnamefont {J.~T.}\ \bibnamefont
			{Kohlhepp}}, \bibinfo {author} {\bibfnamefont {W.~J.~M.}\ \bibnamefont
			{de~Jonge}},\ and\ \bibinfo {author} {\bibfnamefont {B.}~\bibnamefont
			{Koopmans}},\ }\bibfield  {title} {\emph {\bibinfo {title} {Resolving the
				genuine laser-induced ultrafast dynamics of exchange interaction in
				ferromagnet/antiferromagnet bilayers}},\ }\href
	{https://doi.org/10.1103/PhysRevB.81.094435} {\bibfield  {journal} {\bibinfo
			{journal} {Phys. Rev. B}\ }\textbf {\bibinfo {volume} {81}},\ \bibinfo
		{pages} {094435} (\bibinfo {year} {2010})}\BibitemShut {NoStop}%
	\bibitem [{\citenamefont {Tieg}\ \emph {et~al.}(2006)\citenamefont {Tieg},
		\citenamefont {Kuch}, \citenamefont {Wang},\ and\ \citenamefont
		{Kirschner}}]{Tieg.2006}%
	\BibitemOpen
	\bibfield  {author} {\bibinfo {author} {\bibfnamefont {C.}~\bibnamefont
			{Tieg}}, \bibinfo {author} {\bibfnamefont {W.}~\bibnamefont {Kuch}}, \bibinfo
		{author} {\bibfnamefont {S.~G.}\ \bibnamefont {Wang}},\ and\ \bibinfo
		{author} {\bibfnamefont {J.}~\bibnamefont {Kirschner}},\ }\bibfield  {title}
	{\emph {\bibinfo {title} {Growth, structure, and magnetism of
				single-crystalline ${Ni}_{x}{{{M}n}}_{100\ensuremath{-}x}$ films and
				${{N}i}{{Mn}}∕{{C}o}$ bilayers on {C}u(001)}},\ }\href
	{https://doi.org/10.1103/PhysRevB.74.094420} {\bibfield  {journal} {\bibinfo
			{journal} {Phys. Rev. B}\ }\textbf {\bibinfo {volume} {74}},\ \bibinfo
		{pages} {094420} (\bibinfo {year} {2006})}\BibitemShut {NoStop}%
	\bibitem [{\citenamefont {Hagelschuer}\ \emph {et~al.}(2016)\citenamefont
		{Hagelschuer}, \citenamefont {Shokr},\ and\ \citenamefont
		{Kuch}}]{Hagelschuer.2016}%
	\BibitemOpen
	\bibfield  {author} {\bibinfo {author} {\bibfnamefont {T.}~\bibnamefont
			{Hagelschuer}}, \bibinfo {author} {\bibfnamefont {Y.~A.}\ \bibnamefont
			{Shokr}},\ and\ \bibinfo {author} {\bibfnamefont {W.}~\bibnamefont {Kuch}},\
	}\bibfield  {title} {\emph {\bibinfo {title} {Spin-state transition in
				antiferromagnetic ${{{N}i}}_{0.4}{{{M}n}}_{0.6}$ films in {N}i/{N}i{M}n/{N}i
				trilayers on {C}u(001)}},\ }\href
	{https://doi.org/10.1103/PhysRevB.93.054428} {\bibfield  {journal} {\bibinfo
			{journal} {Phys. Rev. B}\ }\textbf {\bibinfo {volume} {93}},\ \bibinfo
		{pages} {054428} (\bibinfo {year} {2016})}\BibitemShut {NoStop}%
	\bibitem [{\citenamefont {Reinhardt}\ \emph {et~al.}(2010)\citenamefont
		{Reinhardt}, \citenamefont {Seifert}, \citenamefont {Busch},\ and\
		\citenamefont {Winter}}]{Reinhardt.2010}%
	\BibitemOpen
	\bibfield  {author} {\bibinfo {author} {\bibfnamefont {M.}~\bibnamefont
			{Reinhardt}}, \bibinfo {author} {\bibfnamefont {J.}~\bibnamefont {Seifert}},
		\bibinfo {author} {\bibfnamefont {M.}~\bibnamefont {Busch}},\ and\ \bibinfo
		{author} {\bibfnamefont {H.}~\bibnamefont {Winter}},\ }\bibfield  {title}
	{\emph {\bibinfo {title} {Magnetic interface coupling between ultrathin {C}o
				and ${{Ni}}_{x}{{Mn}}_{100\ensuremath{-}x}$ films on {C}u(001)}},\ }\href
	{https://doi.org/10.1103/PhysRevB.81.134433} {\bibfield  {journal} {\bibinfo
			{journal} {Phys. Rev. B}\ }\textbf {\bibinfo {volume} {81}},\ \bibinfo
		{pages} {134433} (\bibinfo {year} {2010})}\BibitemShut {NoStop}%
	\bibitem [{\citenamefont {Holldack}\ \emph {et~al.}(2014)\citenamefont
		{Holldack}, \citenamefont {Bahrdt}, \citenamefont {Balzer}, \citenamefont
		{Bovensiepen}, \citenamefont {Brzhezinskaya}, \citenamefont {Erko},
		\citenamefont {Eschenlohr}, \citenamefont {Follath}, \citenamefont {Firsov},
		\citenamefont {Frentrup}, \citenamefont {{Le Guyader}}, \citenamefont
		{Kachel}, \citenamefont {Kuske}, \citenamefont {Mitzner}, \citenamefont
		{M{\"u}ller}, \citenamefont {Pontius}, \citenamefont {Quast}, \citenamefont
		{Radu}, \citenamefont {Schmidt}, \citenamefont {Sch{\"u}ssler-Langeheine},
		\citenamefont {Sperling}, \citenamefont {Stamm}, \citenamefont {Trabant},\
		and\ \citenamefont {F{\"o}hlisch}}]{Holldack.2014}%
	\BibitemOpen
	\bibfield  {author} {\bibinfo {author} {\bibfnamefont {K.}~\bibnamefont
			{Holldack}}, \bibinfo {author} {\bibfnamefont {J.}~\bibnamefont {Bahrdt}},
		\bibinfo {author} {\bibfnamefont {A.}~\bibnamefont {Balzer}}, \bibinfo
		{author} {\bibfnamefont {U.}~\bibnamefont {Bovensiepen}}, \bibinfo {author}
		{\bibfnamefont {M.}~\bibnamefont {Brzhezinskaya}}, \bibinfo {author}
		{\bibfnamefont {A.}~\bibnamefont {Erko}}, \bibinfo {author} {\bibfnamefont
			{A.}~\bibnamefont {Eschenlohr}}, \bibinfo {author} {\bibfnamefont
			{R.}~\bibnamefont {Follath}}, \bibinfo {author} {\bibfnamefont
			{A.}~\bibnamefont {Firsov}}, \bibinfo {author} {\bibfnamefont
			{W.}~\bibnamefont {Frentrup}}, \bibinfo {author} {\bibfnamefont
			{L.}~\bibnamefont {{Le Guyader}}}, \bibinfo {author} {\bibfnamefont
			{T.}~\bibnamefont {Kachel}}, \bibinfo {author} {\bibfnamefont
			{P.}~\bibnamefont {Kuske}}, \bibinfo {author} {\bibfnamefont
			{R.}~\bibnamefont {Mitzner}}, \bibinfo {author} {\bibfnamefont
			{R.}~\bibnamefont {M{\"u}ller}}, \bibinfo {author} {\bibfnamefont
			{N.}~\bibnamefont {Pontius}}, \bibinfo {author} {\bibfnamefont
			{T.}~\bibnamefont {Quast}}, \bibinfo {author} {\bibfnamefont
			{I.}~\bibnamefont {Radu}}, \bibinfo {author} {\bibfnamefont {J.~S.}\
			\bibnamefont {Schmidt}}, \bibinfo {author} {\bibfnamefont {C.}~\bibnamefont
			{Sch{\"u}ssler-Langeheine}}, \bibinfo {author} {\bibfnamefont
			{M.}~\bibnamefont {Sperling}}, \bibinfo {author} {\bibfnamefont
			{C.}~\bibnamefont {Stamm}}, \bibinfo {author} {\bibfnamefont
			{C.}~\bibnamefont {Trabant}},\ and\ \bibinfo {author} {\bibfnamefont
			{A.}~\bibnamefont {F{\"o}hlisch}},\ }\bibfield  {title} {\emph {\bibinfo
			{title} {Femto{S}pe{X}: a versatile optical pump-soft {X}-ray probe facility
				with 100 fs {X}-ray pulses of variable polarization}},\ }\href
	{https://doi.org/10.1107/S1600577514012247} {\bibfield  {journal} {\bibinfo
			{journal} {J. Synchr. Rad.}\ }\textbf {\bibinfo {volume} {21}},\ \bibinfo
		{pages} {1090} (\bibinfo {year} {2014})}\BibitemShut {NoStop}%
	\bibitem [{\citenamefont {Mertins}\ \emph {et~al.}(2002)\citenamefont
		{Mertins}, \citenamefont {Abramsohn}, \citenamefont {Gaupp}, \citenamefont
		{Sch\"afers}, \citenamefont {Gudat}, \citenamefont {Zaharko}, \citenamefont
		{Grimmer},\ and\ \citenamefont {Oppeneer}}]{Mertins.2002}%
	\BibitemOpen
	\bibfield  {author} {\bibinfo {author} {\bibfnamefont {H.-C.}\ \bibnamefont
			{Mertins}}, \bibinfo {author} {\bibfnamefont {D.}~\bibnamefont {Abramsohn}},
		\bibinfo {author} {\bibfnamefont {A.}~\bibnamefont {Gaupp}}, \bibinfo
		{author} {\bibfnamefont {F.}~\bibnamefont {Sch\"afers}}, \bibinfo {author}
		{\bibfnamefont {W.}~\bibnamefont {Gudat}}, \bibinfo {author} {\bibfnamefont
			{O.}~\bibnamefont {Zaharko}}, \bibinfo {author} {\bibfnamefont
			{H.}~\bibnamefont {Grimmer}},\ and\ \bibinfo {author} {\bibfnamefont {P.~M.}\
			\bibnamefont {Oppeneer}},\ }\bibfield  {title} {\emph {\bibinfo {title}
			{Resonant magnetic reflection coefficients at the {F}e $2p$ edge obtained
				with linearly and circularly polarized soft {X}-rays}},\ }\href
	{https://doi.org/10.1103/PhysRevB.66.184404} {\bibfield  {journal} {\bibinfo
			{journal} {Phys. Rev. B}\ }\textbf {\bibinfo {volume} {66}},\ \bibinfo
		{pages} {184404} (\bibinfo {year} {2002})}\BibitemShut {NoStop}%
	\bibitem [{\citenamefont {Schneider}\ \emph {et~al.}(1990)\citenamefont
		{Schneider}, \citenamefont {Bressler}, \citenamefont {Schuster},
		\citenamefont {Kirschner}, \citenamefont {de~Miguel},\ and\ \citenamefont
		{Miranda}}]{Schneider.1990}%
	\BibitemOpen
	\bibfield  {author} {\bibinfo {author} {\bibfnamefont {C.~M.}\ \bibnamefont
			{Schneider}}, \bibinfo {author} {\bibfnamefont {P.}~\bibnamefont {Bressler}},
		\bibinfo {author} {\bibfnamefont {P.}~\bibnamefont {Schuster}}, \bibinfo
		{author} {\bibfnamefont {J.}~\bibnamefont {Kirschner}}, \bibinfo {author}
		{\bibfnamefont {J.~J.}\ \bibnamefont {de~Miguel}},\ and\ \bibinfo {author}
		{\bibfnamefont {R.}~\bibnamefont {Miranda}},\ }\bibfield  {title} {\emph
		{\bibinfo {title} {Curie temperature of ultrathin films of fcc-cobalt
				epitaxially grown on atomically flat {C}u(100) surfaces}},\ }\href
	{https://doi.org/10.1103/PhysRevLett.64.1059} {\bibfield  {journal} {\bibinfo
			{journal} {Phys. Rev. Lett.}\ }\textbf {\bibinfo {volume} {64}},\ \bibinfo
		{pages} {1059} (\bibinfo {year} {1990})}\BibitemShut {NoStop}%
	\bibitem [{\citenamefont {Roth}\ \emph {et~al.}(2012)\citenamefont {Roth},
		\citenamefont {Schellekens}, \citenamefont {Alebrand}, \citenamefont
		{Schmitt}, \citenamefont {Steil}, \citenamefont {Koopmans}, \citenamefont
		{Cinchetti},\ and\ \citenamefont {Aeschlimann}}]{Roth.2012}%
	\BibitemOpen
	\bibfield  {author} {\bibinfo {author} {\bibfnamefont {T.}~\bibnamefont
			{Roth}}, \bibinfo {author} {\bibfnamefont {A.~J.}\ \bibnamefont
			{Schellekens}}, \bibinfo {author} {\bibfnamefont {S.}~\bibnamefont
			{Alebrand}}, \bibinfo {author} {\bibfnamefont {O.}~\bibnamefont {Schmitt}},
		\bibinfo {author} {\bibfnamefont {D.}~\bibnamefont {Steil}}, \bibinfo
		{author} {\bibfnamefont {B.}~\bibnamefont {Koopmans}}, \bibinfo {author}
		{\bibfnamefont {M.}~\bibnamefont {Cinchetti}},\ and\ \bibinfo {author}
		{\bibfnamefont {M.}~\bibnamefont {Aeschlimann}},\ }\bibfield  {title} {\emph
		{\bibinfo {title} {Temperature {D}ependence of {L}aser-{I}nduced
				{D}emagnetization in {N}i: {A} {K}ey for {I}dentifying the {U}nderlying
				{M}echanism}},\ }\href {https://doi.org/10.1103/PhysRevX.2.021006} {\bibfield
		{journal} {\bibinfo  {journal} {Phys. Rev. X}\ }\textbf {\bibinfo {volume}
			{2}},\ \bibinfo {pages} {021006} (\bibinfo {year} {2012})}\BibitemShut
	{NoStop}%
	\bibitem [{\citenamefont {Ghosh}\ \emph {et~al.}(2012)\citenamefont {Ghosh},
		\citenamefont {Auffret}, \citenamefont {Ebels},\ and\ \citenamefont
		{Bailey}}]{Ghosh.2012}%
	\BibitemOpen
	\bibfield  {author} {\bibinfo {author} {\bibfnamefont {A.}~\bibnamefont
			{Ghosh}}, \bibinfo {author} {\bibfnamefont {S.}~\bibnamefont {Auffret}},
		\bibinfo {author} {\bibfnamefont {U.}~\bibnamefont {Ebels}},\ and\ \bibinfo
		{author} {\bibfnamefont {W.~E.}\ \bibnamefont {Bailey}},\ }\bibfield  {title}
	{\emph {\bibinfo {title} {Penetration {D}epth of {T}ransverse {S}pin
				{C}urrent in {U}ltrathin {F}erromagnets}},\ }\href
	{https://doi.org/10.1103/PhysRevLett.109.127202} {\bibfield  {journal}
		{\bibinfo  {journal} {Phys. Rev. Lett.}\ }\textbf {\bibinfo {volume} {109}},\
		\bibinfo {pages} {127202} (\bibinfo {year} {2012})}\BibitemShut {NoStop}%
	\bibitem [{\citenamefont {Alekhin}\ \emph {et~al.}(2017)\citenamefont
		{Alekhin}, \citenamefont {Razdolski}, \citenamefont {Ilin}, \citenamefont
		{Meyburg}, \citenamefont {Diesing}, \citenamefont {Roddatis}, \citenamefont
		{Rungger}, \citenamefont {Stamenova}, \citenamefont {Sanvito}, \citenamefont
		{Bovensiepen},\ and\ \citenamefont {Melnikov}}]{Alekhin.2017}%
	\BibitemOpen
	\bibfield  {author} {\bibinfo {author} {\bibfnamefont {A.}~\bibnamefont
			{Alekhin}}, \bibinfo {author} {\bibfnamefont {I.}~\bibnamefont {Razdolski}},
		\bibinfo {author} {\bibfnamefont {N.}~\bibnamefont {Ilin}}, \bibinfo {author}
		{\bibfnamefont {J.~P.}\ \bibnamefont {Meyburg}}, \bibinfo {author}
		{\bibfnamefont {D.}~\bibnamefont {Diesing}}, \bibinfo {author} {\bibfnamefont
			{V.}~\bibnamefont {Roddatis}}, \bibinfo {author} {\bibfnamefont
			{I.}~\bibnamefont {Rungger}}, \bibinfo {author} {\bibfnamefont
			{M.}~\bibnamefont {Stamenova}}, \bibinfo {author} {\bibfnamefont
			{S.}~\bibnamefont {Sanvito}}, \bibinfo {author} {\bibfnamefont
			{U.}~\bibnamefont {Bovensiepen}},\ and\ \bibinfo {author} {\bibfnamefont
			{A.}~\bibnamefont {Melnikov}},\ }\bibfield  {title} {\emph {\bibinfo {title}
			{Femtosecond {S}pin {C}urrent {P}ulses {G}enerated by the {N}onthermal
				{S}pin-{D}ependent {S}eebeck {E}ffect and {I}nteracting with {F}erromagnets
				in {S}pin {V}alves}},\ }\href
	{https://doi.org/10.1103/PhysRevLett.119.017202} {\bibfield  {journal}
		{\bibinfo  {journal} {Phys. Rev. Lett.}\ }\textbf {\bibinfo {volume} {119}},\
		\bibinfo {pages} {017202} (\bibinfo {year} {2017})}\BibitemShut {NoStop}%
	\bibitem [{\citenamefont {Battiato}\ \emph {et~al.}(2012)\citenamefont
		{Battiato}, \citenamefont {Carva},\ and\ \citenamefont
		{Oppeneer}}]{Battiato.2012}%
	\BibitemOpen
	\bibfield  {author} {\bibinfo {author} {\bibfnamefont {M.}~\bibnamefont
			{Battiato}}, \bibinfo {author} {\bibfnamefont {K.}~\bibnamefont {Carva}},\
		and\ \bibinfo {author} {\bibfnamefont {P.~M.}\ \bibnamefont {Oppeneer}},\
	}\bibfield  {title} {\emph {\bibinfo {title} {Theory of laser-induced
				ultrafast superdiffusive spin transport in layered heterostructures}},\
	}\href {https://doi.org/10.1103/PhysRevB.86.024404} {\bibfield  {journal}
		{\bibinfo  {journal} {Phys. Rev. B}\ }\textbf {\bibinfo {volume} {86}},\
		\bibinfo {pages} {024404} (\bibinfo {year} {2012})}\BibitemShut {NoStop}%
	\bibitem [{\citenamefont {Carpene}\ \emph {et~al.}(2008)\citenamefont
		{Carpene}, \citenamefont {Mancini}, \citenamefont {Dallera}, \citenamefont
		{Brenna}, \citenamefont {Puppin},\ and\ \citenamefont
		{De~Silvestri}}]{Carpene.2008}%
	\BibitemOpen
	\bibfield  {author} {\bibinfo {author} {\bibfnamefont {E.}~\bibnamefont
			{Carpene}}, \bibinfo {author} {\bibfnamefont {E.}~\bibnamefont {Mancini}},
		\bibinfo {author} {\bibfnamefont {C.}~\bibnamefont {Dallera}}, \bibinfo
		{author} {\bibfnamefont {M.}~\bibnamefont {Brenna}}, \bibinfo {author}
		{\bibfnamefont {E.}~\bibnamefont {Puppin}},\ and\ \bibinfo {author}
		{\bibfnamefont {S.}~\bibnamefont {De~Silvestri}},\ }\bibfield  {title} {\emph
		{\bibinfo {title} {Dynamics of electron-magnon interaction and ultrafast
				demagnetization in thin iron films}},\ }\href
	{https://doi.org/10.1103/PhysRevB.78.174422} {\bibfield  {journal} {\bibinfo
			{journal} {Phys. Rev. B}\ }\textbf {\bibinfo {volume} {78}},\ \bibinfo
		{pages} {174422} (\bibinfo {year} {2008})}\BibitemShut {NoStop}%
	\bibitem [{\citenamefont {Eich}\ \emph {et~al.}(2017)\citenamefont {Eich},
		\citenamefont {Pl{\"o}tzing}, \citenamefont {Rollinger}, \citenamefont
		{Emmerich}, \citenamefont {Adam}, \citenamefont {Chen}, \citenamefont
		{Kapteyn}, \citenamefont {Murnane}, \citenamefont {Plucinski}, \citenamefont
		{Steil}, \citenamefont {Stadtm{\"u}ller}, \citenamefont {Cinchetti},
		\citenamefont {Aeschlimann}, \citenamefont {Schneider},\ and\ \citenamefont
		{Mathias}}]{Eich.2017}%
	\BibitemOpen
	\bibfield  {author} {\bibinfo {author} {\bibfnamefont {S.}~\bibnamefont
			{Eich}}, \bibinfo {author} {\bibfnamefont {M.}~\bibnamefont {Pl{\"o}tzing}},
		\bibinfo {author} {\bibfnamefont {M.}~\bibnamefont {Rollinger}}, \bibinfo
		{author} {\bibfnamefont {S.}~\bibnamefont {Emmerich}}, \bibinfo {author}
		{\bibfnamefont {R.}~\bibnamefont {Adam}}, \bibinfo {author} {\bibfnamefont
			{C.}~\bibnamefont {Chen}}, \bibinfo {author} {\bibfnamefont {H.~C.}\
			\bibnamefont {Kapteyn}}, \bibinfo {author} {\bibfnamefont {M.~M.}\
			\bibnamefont {Murnane}}, \bibinfo {author} {\bibfnamefont {L.}~\bibnamefont
			{Plucinski}}, \bibinfo {author} {\bibfnamefont {D.}~\bibnamefont {Steil}},
		\bibinfo {author} {\bibfnamefont {B.}~\bibnamefont {Stadtm{\"u}ller}},
		\bibinfo {author} {\bibfnamefont {M.}~\bibnamefont {Cinchetti}}, \bibinfo
		{author} {\bibfnamefont {M.}~\bibnamefont {Aeschlimann}}, \bibinfo {author}
		{\bibfnamefont {C.~M.}\ \bibnamefont {Schneider}},\ and\ \bibinfo {author}
		{\bibfnamefont {S.}~\bibnamefont {Mathias}},\ }\bibfield  {title} {\emph
		{\bibinfo {title} {Band structure evolution during the ultrafast
				ferromagnetic-paramagnetic phase transition in cobalt}},\ }\href
	{https://doi.org/10.1126/sciadv.1602094} {\bibfield  {journal} {\bibinfo
			{journal} {Sci. Adv.}\ }\textbf {\bibinfo {volume} {3}},\ \bibinfo {pages}
		{e1602094} (\bibinfo {year} {2017})}\BibitemShut {NoStop}%
	\bibitem [{\citenamefont {Gort}\ \emph {et~al.}(2018)\citenamefont {Gort},
		\citenamefont {B{\"u}hlmann}, \citenamefont {D{\"a}ster}, \citenamefont
		{Salvatella}, \citenamefont {Hartmann}, \citenamefont {Zemp}, \citenamefont
		{Holenstein}, \citenamefont {Stieger}, \citenamefont {Fognini}, \citenamefont
		{Michlmayr}, \citenamefont {B{\"a}hler}, \citenamefont {Vaterlaus},\ and\
		\citenamefont {Acremann}}]{Gort.2018}%
	\BibitemOpen
	\bibfield  {author} {\bibinfo {author} {\bibfnamefont {R.}~\bibnamefont
			{Gort}}, \bibinfo {author} {\bibfnamefont {K.}~\bibnamefont {B{\"u}hlmann}},
		\bibinfo {author} {\bibfnamefont {S.}~\bibnamefont {D{\"a}ster}}, \bibinfo
		{author} {\bibfnamefont {G.}~\bibnamefont {Salvatella}}, \bibinfo {author}
		{\bibfnamefont {N.}~\bibnamefont {Hartmann}}, \bibinfo {author}
		{\bibfnamefont {Y.}~\bibnamefont {Zemp}}, \bibinfo {author} {\bibfnamefont
			{S.}~\bibnamefont {Holenstein}}, \bibinfo {author} {\bibfnamefont
			{C.}~\bibnamefont {Stieger}}, \bibinfo {author} {\bibfnamefont
			{A.}~\bibnamefont {Fognini}}, \bibinfo {author} {\bibfnamefont {T.~U.}\
			\bibnamefont {Michlmayr}}, \bibinfo {author} {\bibfnamefont {T.}~\bibnamefont
			{B{\"a}hler}}, \bibinfo {author} {\bibfnamefont {A.}~\bibnamefont
			{Vaterlaus}},\ and\ \bibinfo {author} {\bibfnamefont {Y.}~\bibnamefont
			{Acremann}},\ }\bibfield  {title} {\emph {\bibinfo {title} {Early {S}tages of
				{U}ltrafast {S}pin {D}ynamics in a 3d {F}erromagnet}},\ }\href
	{https://doi.org/10.1103/PhysRevLett.121.087206} {\bibfield  {journal}
		{\bibinfo  {journal} {Phys. Rev. Lett.}\ }\textbf {\bibinfo {volume} {121}},\
		\bibinfo {pages} {087206} (\bibinfo {year} {2018})}\BibitemShut {NoStop}%
	\bibitem [{\citenamefont {Schellekens}\ \emph {et~al.}(2013)\citenamefont
		{Schellekens}, \citenamefont {Verhoeven}, \citenamefont {Vader},\ and\
		\citenamefont {Koopmans}}]{Schellekens.2013}%
	\BibitemOpen
	\bibfield  {author} {\bibinfo {author} {\bibfnamefont {A.~J.}\ \bibnamefont
			{Schellekens}}, \bibinfo {author} {\bibfnamefont {W.}~\bibnamefont
			{Verhoeven}}, \bibinfo {author} {\bibfnamefont {T.~N.}\ \bibnamefont
			{Vader}},\ and\ \bibinfo {author} {\bibfnamefont {B.}~\bibnamefont
			{Koopmans}},\ }\bibfield  {title} {\emph {\bibinfo {title} {Investigating the
				contribution of superdiffusive transport to ultrafast demagnetization of
				ferromagnetic thin films}},\ }\href {https://doi.org/10.1063/1.4812658}
	{\bibfield  {journal} {\bibinfo  {journal} {Appl. Phys. Lett.}\ }\textbf
		{\bibinfo {volume} {102}},\ \bibinfo {pages} {252408} (\bibinfo {year}
		{2013})}\BibitemShut {NoStop}%
	\bibitem [{\citenamefont {Wieczorek}\ \emph {et~al.}(2015)\citenamefont
		{Wieczorek}, \citenamefont {Eschenlohr}, \citenamefont {Weidtmann},
		\citenamefont {R\"osner}, \citenamefont {Bergeard}, \citenamefont
		{Tarasevitch}, \citenamefont {Wehling},\ and\ \citenamefont
		{Bovensiepen}}]{Wieczorek.2015}%
	\BibitemOpen
	\bibfield  {author} {\bibinfo {author} {\bibfnamefont {J.}~\bibnamefont
			{Wieczorek}}, \bibinfo {author} {\bibfnamefont {A.}~\bibnamefont
			{Eschenlohr}}, \bibinfo {author} {\bibfnamefont {B.}~\bibnamefont
			{Weidtmann}}, \bibinfo {author} {\bibfnamefont {M.}~\bibnamefont {R\"osner}},
		\bibinfo {author} {\bibfnamefont {N.}~\bibnamefont {Bergeard}}, \bibinfo
		{author} {\bibfnamefont {A.}~\bibnamefont {Tarasevitch}}, \bibinfo {author}
		{\bibfnamefont {T.~O.}\ \bibnamefont {Wehling}},\ and\ \bibinfo {author}
		{\bibfnamefont {U.}~\bibnamefont {Bovensiepen}},\ }\bibfield  {title} {\emph
		{\bibinfo {title} {Separation of ultrafast spin currents and spin-flip
				scattering in {C}o/{C}u(001) driven by femtosecond laser excitation employing
				the complex magneto-optical {K}err effect}},\ }\href
	{https://doi.org/10.1103/PhysRevB.92.174410} {\bibfield  {journal} {\bibinfo
			{journal} {Phys. Rev. B}\ }\textbf {\bibinfo {volume} {92}},\ \bibinfo
		{pages} {174410} (\bibinfo {year} {2015})}\BibitemShut {NoStop}%
	\bibitem [{\citenamefont {Carva}\ \emph {et~al.}(2011)\citenamefont {Carva},
		\citenamefont {Battiato},\ and\ \citenamefont {Oppeneer}}]{Carva.2011}%
	\BibitemOpen
	\bibfield  {author} {\bibinfo {author} {\bibfnamefont {K.}~\bibnamefont
			{Carva}}, \bibinfo {author} {\bibfnamefont {M.}~\bibnamefont {Battiato}},\
		and\ \bibinfo {author} {\bibfnamefont {P.~M.}\ \bibnamefont {Oppeneer}},\
	}\bibfield  {title} {\emph {\bibinfo {title} {Ab {I}nitio {I}nvestigation of
				the {E}lliott-{Y}afet {E}lectron-{P}honon {M}echanism in {L}aser-{I}nduced
				{U}ltrafast {D}emagnetization}},\ }\href
	{https://doi.org/10.1103/PhysRevLett.107.207201} {\bibfield  {journal}
		{\bibinfo  {journal} {Phys. Rev. Lett.}\ }\textbf {\bibinfo {volume} {107}},\
		\bibinfo {pages} {207201} (\bibinfo {year} {2011})}\BibitemShut {NoStop}%
	\bibitem [{\citenamefont {Essert}\ and\ \citenamefont
		{Schneider}(2011)}]{Essert.2011}%
	\BibitemOpen
	\bibfield  {author} {\bibinfo {author} {\bibfnamefont {S.}~\bibnamefont
			{Essert}}\ and\ \bibinfo {author} {\bibfnamefont {H.~C.}\ \bibnamefont
			{Schneider}},\ }\bibfield  {title} {\emph {\bibinfo {title} {Electron-phonon
				scattering dynamics in ferromagnetic metals and their influence on ultrafast
				demagnetization processes}},\ }\href
	{https://doi.org/10.1103/PhysRevB.84.224405} {\bibfield  {journal} {\bibinfo
			{journal} {Phys. Rev. B}\ }\textbf {\bibinfo {volume} {84}},\ \bibinfo
		{pages} {224405} (\bibinfo {year} {2011})}\BibitemShut {NoStop}%
	\bibitem [{\citenamefont {Schellekens}\ and\ \citenamefont
		{Koopmans}(2013)}]{Schellekens.2013b}%
	\BibitemOpen
	\bibfield  {author} {\bibinfo {author} {\bibfnamefont {A.~J.}\ \bibnamefont
			{Schellekens}}\ and\ \bibinfo {author} {\bibfnamefont {B.}~\bibnamefont
			{Koopmans}},\ }\bibfield  {title} {\emph {\bibinfo {title} {Comparing
				ultrafast demagnetization rates between competing models for finite
				temperature magnetism}},\ }\href
	{https://doi.org/10.1103/PhysRevLett.110.217204} {\bibfield  {journal}
		{\bibinfo  {journal} {Phys. Rev. Lett.}\ }\textbf {\bibinfo {volume} {110}},\
		\bibinfo {pages} {217204} (\bibinfo {year} {2013})}\BibitemShut {NoStop}%
	\bibitem [{\citenamefont {Mueller}\ \emph {et~al.}(2013)\citenamefont
		{Mueller}, \citenamefont {Baral}, \citenamefont {Vollmar}, \citenamefont
		{Cinchetti}, \citenamefont {Aeschlimann}, \citenamefont {Schneider},\ and\
		\citenamefont {Rethfeld}}]{Mueller.2013}%
	\BibitemOpen
	\bibfield  {author} {\bibinfo {author} {\bibfnamefont {B.~Y.}\ \bibnamefont
			{Mueller}}, \bibinfo {author} {\bibfnamefont {A.}~\bibnamefont {Baral}},
		\bibinfo {author} {\bibfnamefont {S.}~\bibnamefont {Vollmar}}, \bibinfo
		{author} {\bibfnamefont {M.}~\bibnamefont {Cinchetti}}, \bibinfo {author}
		{\bibfnamefont {M.}~\bibnamefont {Aeschlimann}}, \bibinfo {author}
		{\bibfnamefont {H.~C.}\ \bibnamefont {Schneider}},\ and\ \bibinfo {author}
		{\bibfnamefont {B.}~\bibnamefont {Rethfeld}},\ }\bibfield  {title} {\emph
		{\bibinfo {title} {Feedback {E}ffect during {U}ltrafast {D}emagnetization
				{D}ynamics in {F}erromagnets}},\ }\href
	{https://doi.org/10.1103/PhysRevLett.111.167204} {\bibfield  {journal}
		{\bibinfo  {journal} {Phys. Rev. Lett.}\ }\textbf {\bibinfo {volume} {111}},\
		\bibinfo {pages} {167204} (\bibinfo {year} {2013})}\BibitemShut {NoStop}%
	\bibitem [{\citenamefont {Choi}\ \emph {et~al.}(2014)\citenamefont {Choi},
		\citenamefont {Min}, \citenamefont {Lee},\ and\ \citenamefont
		{Cahill}}]{Choi.2014}%
	\BibitemOpen
	\bibfield  {author} {\bibinfo {author} {\bibfnamefont {G.-M.}\ \bibnamefont
			{Choi}}, \bibinfo {author} {\bibfnamefont {B.-C.}\ \bibnamefont {Min}},
		\bibinfo {author} {\bibfnamefont {K.-J.}\ \bibnamefont {Lee}},\ and\ \bibinfo
		{author} {\bibfnamefont {D.~G.}\ \bibnamefont {Cahill}},\ }\bibfield  {title}
	{\emph {\bibinfo {title} {{Spin current generated by thermally driven
					ultrafast demagnetization}}},\ }\href {https://doi.org/10.1038/ncomms5334}
	{\bibfield  {journal} {\bibinfo  {journal} {Nat. Commun.}\ }\textbf {\bibinfo
			{volume} {5}},\ \bibinfo {pages} {4334} (\bibinfo {year} {2014})}\BibitemShut
	{NoStop}%
	\bibitem [{\citenamefont {Choi}\ and\ \citenamefont
		{Cahill}(2014)}]{Choi.2014b}%
	\BibitemOpen
	\bibfield  {author} {\bibinfo {author} {\bibfnamefont {G.-M.}\ \bibnamefont
			{Choi}}\ and\ \bibinfo {author} {\bibfnamefont {D.~G.}\ \bibnamefont
			{Cahill}},\ }\bibfield  {title} {\emph {\bibinfo {title} {Kerr rotation in
				{C}u, {A}g, and {A}u driven by spin accumulation and spin-orbit coupling}},\
	}\href {https://doi.org/10.1103/PhysRevB.90.214432} {\bibfield  {journal}
		{\bibinfo  {journal} {Phys. Rev. B}\ }\textbf {\bibinfo {volume} {90}},\
		\bibinfo {pages} {214432} (\bibinfo {year} {2014})}\BibitemShut {NoStop}%
	\bibitem [{\citenamefont {Wu}\ \emph {et~al.}(2018)\citenamefont {Wu},
		\citenamefont {Liu},\ and\ \citenamefont {Luo}}]{Wu.2018}%
	\BibitemOpen
	\bibfield  {author} {\bibinfo {author} {\bibfnamefont {X.}~\bibnamefont
			{Wu}}, \bibinfo {author} {\bibfnamefont {Z.}~\bibnamefont {Liu}},\ and\
		\bibinfo {author} {\bibfnamefont {T.}~\bibnamefont {Luo}},\ }\bibfield
	{title} {\emph {\bibinfo {title} {Magnon and phonon dispersion, lifetime, and
				thermal conductivity of iron from spin-lattice dynamics simulations}},\
	}\href {https://doi.org/10.1063/1.5020611} {\bibfield  {journal} {\bibinfo
			{journal} {J. Appl. Phys.}\ }\textbf {\bibinfo {volume} {123}},\ \bibinfo
		{pages} {085109} (\bibinfo {year} {2018})}\BibitemShut {NoStop}%
	\bibitem [{\citenamefont {Wadley}\ \emph {et~al.}(2016)\citenamefont {Wadley},
		\citenamefont {Howells}, \citenamefont {\v{Z}elezn\'y}, \citenamefont
		{Andrews}, \citenamefont {Hills}, \citenamefont {Campion}, \citenamefont
		{Nov\'ak}, \citenamefont {Olejn\'ik}, \citenamefont {Maccherozzi},
		\citenamefont {Dhesi}, \citenamefont {Martin}, \citenamefont {Wagner},
		\citenamefont {Wunderlich}, \citenamefont {Freimuth}, \citenamefont
		{Mokrousov}, \citenamefont {Kune\v{s}}, \citenamefont {Chauhan},
		\citenamefont {Grzybowski}, \citenamefont {Rushforth}, \citenamefont
		{Edmonds}, \citenamefont {Gallagher},\ and\ \citenamefont
		{Jungwirth}}]{Wadley.2016}%
	\BibitemOpen
	\bibfield  {author} {\bibinfo {author} {\bibfnamefont {P.}~\bibnamefont
			{Wadley}}, \bibinfo {author} {\bibfnamefont {B.}~\bibnamefont {Howells}},
		\bibinfo {author} {\bibfnamefont {J.}~\bibnamefont {\v{Z}elezn\'y}}, \bibinfo
		{author} {\bibfnamefont {C.}~\bibnamefont {Andrews}}, \bibinfo {author}
		{\bibfnamefont {V.}~\bibnamefont {Hills}}, \bibinfo {author} {\bibfnamefont
			{R.~P.}\ \bibnamefont {Campion}}, \bibinfo {author} {\bibfnamefont
			{V.}~\bibnamefont {Nov\'ak}}, \bibinfo {author} {\bibfnamefont
			{K.}~\bibnamefont {Olejn\'ik}}, \bibinfo {author} {\bibfnamefont
			{F.}~\bibnamefont {Maccherozzi}}, \bibinfo {author} {\bibfnamefont {S.~S.}\
			\bibnamefont {Dhesi}}, \bibinfo {author} {\bibfnamefont {S.~Y.}\ \bibnamefont
			{Martin}}, \bibinfo {author} {\bibfnamefont {T.}~\bibnamefont {Wagner}},
		\bibinfo {author} {\bibfnamefont {J.}~\bibnamefont {Wunderlich}}, \bibinfo
		{author} {\bibfnamefont {F.}~\bibnamefont {Freimuth}}, \bibinfo {author}
		{\bibfnamefont {Y.}~\bibnamefont {Mokrousov}}, \bibinfo {author}
		{\bibfnamefont {J.}~\bibnamefont {Kune\v{s}}}, \bibinfo {author}
		{\bibfnamefont {J.~S.}\ \bibnamefont {Chauhan}}, \bibinfo {author}
		{\bibfnamefont {M.~J.}\ \bibnamefont {Grzybowski}}, \bibinfo {author}
		{\bibfnamefont {A.~W.}\ \bibnamefont {Rushforth}}, \bibinfo {author}
		{\bibfnamefont {K.~W.}\ \bibnamefont {Edmonds}}, \bibinfo {author}
		{\bibfnamefont {B.~L.}\ \bibnamefont {Gallagher}},\ and\ \bibinfo {author}
		{\bibfnamefont {T.}~\bibnamefont {Jungwirth}},\ }\bibfield  {title} {\emph
		{\bibinfo {title} {Electrical switching of an antiferromagnet}},\ }\href
	{https://doi.org/10.1126/science.aab1031} {\bibfield  {journal} {\bibinfo
			{journal} {Science}\ }\textbf {\bibinfo {volume} {351}},\ \bibinfo {pages}
		{587} (\bibinfo {year} {2016})}\BibitemShut {NoStop}%
	\bibitem [{\citenamefont {Ohta}\ and\ \citenamefont {Ishida}(1990)}]{Ohta.90}%
	\BibitemOpen
	\bibfield  {author} {\bibinfo {author} {\bibfnamefont {K.}~\bibnamefont
			{Ohta}}\ and\ \bibinfo {author} {\bibfnamefont {H.}~\bibnamefont {Ishida}},\
	}\bibfield  {title} {\emph {\bibinfo {title} {Matrix formalism for
				calculation of electric field intensity of light in stratified multilayered
				films}},\ }\href {https://doi.org/10.1364/AO.29.001952} {\bibfield  {journal}
		{\bibinfo  {journal} {Appl. Opt.}\ }\textbf {\bibinfo {volume} {29}},\
		\bibinfo {pages} {1952} (\bibinfo {year} {1990})}\BibitemShut {NoStop}%
	\bibitem [{\citenamefont {Johnson}\ and\ \citenamefont
		{Christy}(1974)}]{Johnson.1974}%
	\BibitemOpen
	\bibfield  {author} {\bibinfo {author} {\bibfnamefont {P.~B.}\ \bibnamefont
			{Johnson}}\ and\ \bibinfo {author} {\bibfnamefont {R.~W.}\ \bibnamefont
			{Christy}},\ }\bibfield  {title} {\emph {\bibinfo {title} {Optical constants
				of transition metals: {T}i, {V}, {C}r, {M}n, {F}e, {C}o, {N}i, and {P}d}},\
	}\href {https://doi.org/10.1103/PhysRevB.9.5056} {\bibfield  {journal}
		{\bibinfo  {journal} {Phys. Rev. B}\ }\textbf {\bibinfo {volume} {9}},\
		\bibinfo {pages} {5056} (\bibinfo {year} {1974})}\BibitemShut {NoStop}%
	\bibitem [{\citenamefont {Cerd\'a}\ \emph {et~al.}(1993)\citenamefont
		{Cerd\'a}, \citenamefont {de~Andres}, \citenamefont {Cebollada},
		\citenamefont {Miranda}, \citenamefont {Navas}, \citenamefont {Schuster},
		\citenamefont {Schneider},\ and\ \citenamefont {Kirschner}}]{Cerda.1993}%
	\BibitemOpen
	\bibfield  {author} {\bibinfo {author} {\bibfnamefont {J.~R.}\ \bibnamefont
			{Cerd\'a}}, \bibinfo {author} {\bibfnamefont {P.~L.}\ \bibnamefont
			{de~Andres}}, \bibinfo {author} {\bibfnamefont {A.}~\bibnamefont
			{Cebollada}}, \bibinfo {author} {\bibfnamefont {R.}~\bibnamefont {Miranda}},
		\bibinfo {author} {\bibfnamefont {E.}~\bibnamefont {Navas}}, \bibinfo
		{author} {\bibfnamefont {P.}~\bibnamefont {Schuster}}, \bibinfo {author}
		{\bibfnamefont {C.~M.}\ \bibnamefont {Schneider}},\ and\ \bibinfo {author}
		{\bibfnamefont {J.}~\bibnamefont {Kirschner}},\ }\bibfield  {title} {\emph
		{\bibinfo {title} {Epitaxial growth of cobalt films on {C}u(100): a
				crystallographic {LEED} determination}},\ }\href
	{https://doi.org/10.1088/0953-8984/5/14/005} {\bibfield  {journal} {\bibinfo
			{journal} {J. Phys.: Condens. Matt.}\ }\textbf {\bibinfo {volume} {5}},\
		\bibinfo {pages} {2055} (\bibinfo {year} {1993})}\BibitemShut {NoStop}%
\end{thebibliography}
\bibliographystyle{apsrev4-1}

\end{document}